\documentclass[prx, twocolumn, english, amsmath,amssymb, tabularx,
 superscriptaddress, floatfix, longbibliography]{revtex4-2}

\usepackage{xfrac}
\usepackage{float}
\usepackage{graphicx}
\usepackage{dsfont,mathtools}
\usepackage{dcolumn}
\usepackage{bm}
\usepackage{braket}
\usepackage{lineno}
\usepackage{array}
\usepackage{tabularx}
\usepackage{multirow}
\usepackage{xcolor}
\usepackage{ulem}
\usepackage{tablefootnote}
 \usepackage{tikz}
 \usetikzlibrary{shapes.geometric, arrows}
 \usepackage{tkz-euclide}
\usepackage{qcircuit}
\usepackage{comment}
\usepackage{hyperref}
\hypersetup{
    colorlinks=true,       
    linkcolor=cyan,          
    citecolor=magenta,        
    filecolor=magenta,      
    urlcolor=cyan,           
    runcolor=cyan
}

\begin{document}

\title{Exploiting Maximally Mixed States for Spectral Estimation by Time Evolution}

\author{Kaelyn J. Ferris}
\affiliation{IBM Quantum, IBM T.J. Watson Research Center, Yorktown Heights, New York 10598, USA}

\author{Zihang Wang}
\affiliation{Department of Physics, University of California Santa Barbara, Santa Barbara, California 93106, USA}

\author{Itay Hen}
\affiliation{Information Sciences Institute, University of Southern California, Marina del Rey, CA 90292, USA}
\affiliation{Department of Physics and Astronomy, and Center for Quantum Information Science \& Technology, University of Southern California, Los Angeles, California 90089, USA}

\author{Amir Kalev}
\affiliation{Information Sciences Institute, University of Southern California, Arlington, VA 22203, USA}
\affiliation{Department of Physics and Astronomy, University of Southern California, Los Angeles, California 90089, USA}

\author{Nicholas T. Bronn}
\affiliation{IBM Quantum, IBM T.J. Watson Research Center, Yorktown Heights, New York 10598, USA}

\author{Vojt\v{e}ch Vl\v{c}ek}
\affiliation{Department of Chemistry and Biochemistry, University of California, Santa Barbara, CA 93106, USA}
\affiliation{Department of Materials, University of California, Santa Barbara, CA 93106, USA}

\date{\today}

\begin{abstract}
\noindent We introduce a novel approach for estimating the spectrum of quantum many-body Hamiltonians, and more generally, of Hermitian operators, using quantum time evolution. In our approach we are evolving a maximally mixed state under the Hamiltonian of interest and collecting specific time-series measurements to estimate its spectrum. We demonstrate the advantage of our technique over currently used classical statistical sampling methods. We showcase our approach by experimentally estimating the spectral decomposition of a 2-qubit Heisenberg Hamiltonian on an IBM Quantum backend. For this purpose, we develop a hardware-efficient decomposition that controls $n$-qubit Pauli rotations against the physically closest qubit alongside expressing two-qubit rotations in terms of the native entangling interaction. This substantially reduced the accumulation of errors from noisy two-qubit operations in time evolution simulation protocols. We conclude by discussing the potential impact of our work and the future directions of research it opens.  
\end{abstract}

\maketitle

\section{Introduction}

Tackling the quantum many-body problem for any realistic system is, besides fundamental theoretical advances, inevitably intertwined with the developments of new computational techniques.\cite{martin_2016} The problem complexity scales exponentially with the system size. Yet, the physically interesting phenomena are typically related to finding the ground or selected excited states (i.e., those accessible on characteristic energy scales), the relevant observables, and measurable correlation functions. In this context, a class of random sampling methods has a key advantage over deterministic calculations as they allow to offset the steep computational cost by recasting the desired quantities (e.g., correlation functions) as statistical estimators over random realizations that explore the Hilbert space of the problem~\cite{StochasticReview}. The most conceptually straightforward problem is associated with determining the spectrum of a Hamiltonian, $H$, by the time evolution of random vectors and the spectral analysis of their autocorrelation. Since only information on selected states is typically desired, this may be achieved by employing a set of random states in a particular subspace.

In principle,  a single vector constructed as a uniformly weighted linear combination of the selected stationary states provides access to the spectrum of interest. However, as the spectral decomposition of $H$ these stationary states are not known \textit{a priori}. As a result, each realization represents a  particular sampling of the overlap with the eigenvectors of $H$. Since every sample has varying (and non-uniform) overlap, the calculation thus needs to be repeated and averaged, while the standard deviation of the spectrum estimation decreases as $1/\sqrt{N}$ with the number of samplings $N$. In principle, the exact results are guaranteed (for unbiased sampling) when $N\to  \infty$. The prefactor of the statistical error is largely determined by the ``information redundancy’’ in the sampled subspace. Random sampling techniques for weakly correlated systems generally exhibit self-averaging and the desired quantity is well captured with only a few random vectors leading to a significant speedup of perturbative many-body techniques.\cite{RoiDannyGW,VojtechStochGW,vlcek2018swift,StochasticVertexcorrections} For strongly correlated systems, however, the problem representation cannot be simply reduced and converging the statistical estimator requires $N$ similar to the dimensionality of the correlated Hilbert space. 

In contrast to classical random sampling methods, quantum mechanics allows the realization of uniformly distributed random vectors sampling the Hilbert (sub)space in the form of a {\it single} maximally mixed state. As such, each such random vector has uniform overlap with the eigenvectors of $H$.  Manipulating in parallel all the vectors that span the relevant subspace on a quantum computer may  provide a distinct advantage over a classical counterpart. Namely, if the system is highly degenerate, a random sampling of pure states may introduce bias in the evaluated multiplicity of the power spectrum which can be avoided by sampling directly from a maximally mixed sate.  For each preparation, the quantum phase estimation (QPE) algorithm~\cite{Abrams1997, Lloyd1996} allows one in principle to calculate a single eigenvalue of $H$ with uniform probability. This process can be repeated until all $N$ eigenvalues have been determined. The quantum method therefore does not suffer from the overhead of classical random sampling, which becomes particularly demanding as applied to strongly-correlated systems. However, QPE is not viable on near term noisy quantum computers.  As such, several modifications to this protocol have been proposed such as robust phase estimation~\cite{kimmel-rpe, kenneth-rpe}, variational quantum phase estimation~\cite{yizhi-vqpe-krylov, klymko-vqpe,shen-vqpe, cohn-qfd}, exploiting spectrographic techniques~\cite{Stenger2022}, or by utilizing mid-circuit measurements~\cite{Corcoles2021}.

In this work, we outline and practically demonstrate an approach to eigenvalue spectrum estimation -- based on a time-series analysis of the Hadamard test~\cite{Somma2002, Somma2019} -- in which a truly random (uniformly distributed) vector is prepared from a maximally mixed state on a quantum computer. Here, the states span the entire Hilbert space, but its projection on a selected subspace can be easily performed by filtering~\cite{StochasticReview}.
We further demonstrate the advantage of the quantum representation over the classical (statistical) sampling methods as well as how the two approaches scale with the system size. We further provide a detailed discussion of the practical limitations of these methods.  

In Section~\ref{section:max-mixed} we generally describe the stochastic and quantum approaches to approximating the spectrum of a given Hamiltonian by time evolution.  In Section~\ref{section:spectrum} we discuss a simple example system on which to benchmark these two approaches to the eigenvalue spectrum estimation.  The results of these approaches are then discussed in Section~\ref{section:results} and conclusions and future perspectives are given in Section~\ref{section:conclusions}.

\section{Spectral Estimation via Time Evolution}\label{section:max-mixed}
Let $H$ be a time-independent Hamiltonian on  Hilbert space of dimension $d$. The time evolution operator under this Hamiltonian is given by
\begin{align}
U(t)=\mathrm{e}^{-i H t} \quad {\rm taking} \quad \ket{\psi(0)} \xlongrightarrow{U} \ket{\psi(t)}
\end{align}
where $t$ denotes the evolution time and $\psi$ is an arbitrary (pure) quantum state ($\hbar \equiv 1$ throughout). A simple yet potentially powerful realization is that the trace of $U(t)$, which we denote by $u(t)$, has information about the entire spectrum of $H$
\begin{align}
    u(t)=\text{Tr}\big(U(t)\big)=\sum_j \text{e}^{-i\lambda_j t},
    \label{max mix state}
\end{align}
where the $\lambda_j$ are the eigenvalues of the Hamiltonian (which may be degenerate). Consequently, the Fourier transform of  $u(t)$, $\tilde{u}(\omega)={\cal F}(u(t))$, provides the eigenvalue spectrum of $H$ (including the multiplicity of degenerate eigenvalues) in the frequency domain: 
\begin{align}
    u(t)\xlongrightarrow{{\cal F}}  \tilde{u}(\omega)=2\pi\sum_j \delta(\omega-\lambda_j).
\end{align}
in the limit of infinite time.  
This mathematical setup inspires numerous techniques~\cite{HutchinsonTrace,NeuhauserFD,NeuhauserFD2,Mandelshtam}  and it is also employed in the proposed quantum protocol for estimating the spectrum of $H$. In the first case, it is employed in a suite of \textit{random sampling algorithms} that emerged as numerically advantageous in evaluating expectation values and correlators~\cite{StochasticReview}. A classical implementation of stochastic sampling of the spectrum of the system Hamiltonian employs a set of random vectors $\{\ket{\psi_{\rm stoc}^{(k)}}\}_k$ (each representing a pure quantum state) for which the stochastic resolution of identity holds: $\lim_{K \to \infty} \frac{1}{K}\sum_{k=1}^K \ket{\psi^{(k)}_{\rm stoc}}\bra{\psi^{(k)}_{\rm stoc}} = {\mathbf I}$. In practice, the number of stochastic vectors, $K$, is finite, and each sampling state is generated as a linear combination:
\begin{equation}
\ket{\psi_{\rm stoc}^{(k)}} = \frac{1}{\sqrt{S}}\sum_{j=1}^S \alpha_j^{(k)} \ket{E_j^{(k)}},
\label{Stochastic state}
\end{equation}
where $k\in[1,K]$ labels the state, $\ket{E_j^{(k)}}$ are efficiently-computable basis states, typically chosen as eigenstates of an auxiliary Hamiltonian $\tilde{H}$. Here, $\alpha_j^{(k)}\equiv e^{i\theta_j^{(k)}}$ is   
a random phase; the choice of $\tilde{H}$ is not unique, and, typically, one employs $\tilde{H}$ that is related to the problem in question, e.g., the corresponding mean-field approximation to $H$ is assumed   in the context of perturbation expansion \cite{vlcek2018swift,StochasticVertexcorrections}. The corresponding spectrum for each random state has the form, $\tilde\psi_{\rm stoc}^{(k)}(\omega) =\mathcal{F}\left(\psi_{\rm stoc}^{(k)}(t)\right)$ where
\begin{align}
\psi_{\rm stoc}^{(k)}(t)=\bra{\psi_{\rm stoc}^{(k)}} e^{-i Ht} \ket{\psi_{\rm stoc}^{(k)}}.
\end{align}
The statistical average of the sample spectrum should converge, with a variance scaling $\sigma^2 \sim 1/K^2$, to the  spectrum of $H$,
\begin{equation}
    \tilde{\psi}_{K}(\omega)=\frac{1}{K}\sum_{k=1}^{K} \tilde\psi_{\rm stoc}^{(k)}(\omega), \hspace{0.2cm}  \lim_{K\to \infty} \tilde{\psi}_{K}(\omega)=  \tilde{u}(\omega),
\end{equation}
where the last equality uses the central limit theorem.  

In practice, the evolution operator, truncated to first order, has the form
\begin{equation}
   U(T)= \mathrm{exp}\left(-i H T\right) \approx \left(\mathbb{I}-i H \Delta t\right)^n, 
   \label{truncation}
\end{equation}
where $\Delta t=T/n$ defines the unit time step, and the total evolution time $T$ is inversely proportional to the frequency resolution, $\Delta\omega= 2\pi/T$. The time resolution must be small compared to the inverse of the maximum eigenvalue ($\Delta t \ll 1/\omega_{\rm max}$) in order to perform the truncation in Eq.~\ref{truncation}. Additionally, the frequency resolution $\Delta \omega \ll \mathrm{Min}|\omega_{i}-\omega_j|$ must be small enough to resolve the differences between eigenvalues. This sets the minimum total evolution time: $T \gg 2\pi/\mathrm{Min}|\omega_{i}-\omega_j|$. The ratio between these two time scales $T/\Delta t$ sets the operation costs $\epsilon \sim \mathcal{O}(2\pi \omega_{\rm max}/\mathrm{Min}|\omega_{i}-\omega_j|)$. As the system size grows, the minimal difference between eigenvalues generally reduces, and the maximum eigenvalue generally increases, resulting in an increased computational cost.

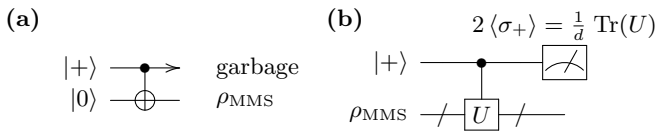
\begin{figure}[t!]
\begin{flushleft}
{\textbf{(a)}}~
\hspace{0.6cm}
\Qcircuit @C=1em @R=0.8em {
 & & & & \\
 & & & & \\
  \lstick{\ket{+}} & \ctrl{1} & \qwa& \rstick{\rm garbage}\\
  \lstick{\ket{0}}  & \targ  &\qw& \rstick{\rho_{\rm MMS}}
} \qquad \qquad {\textbf{(b)}} \qquad
\Qcircuit @C=0.9em @R=0.8em {
  & & & & \mbox{$2\left<\sigma_+\right>=\frac1{d}\;\text{Tr}(U)$} & \\
  \lstick{\ket{+}} & \qw & \ctrl{1} & \qw & \meter & \\
  \lstick{\rho_{\rm MMS}} & {/} \qw & \gate{U} & {/}  \qw & \qw
  }
\end{flushleft}
\caption{{\bf An illustration of a protocol to measure $\text{Tr}(U)$.} (a)~Circuit diagram generating the maximally mixed state for a single qubit. Two qubits are first prepared in a Bell state, then one is discarded while the other is operated on further. (b) A circuit for measuring $\text{Tr}(U)$, where $\rho_{\rm MMS}=\frac1{d}\mathbf{I}$. The circuit is repeated many times. In each realization we measure the pointer qubit (top register in the figure) in either the Pauli-$X$ basis or the Pauli-$Y$ basis (the basis of measurement is implicit in the meter). Over many repetitions we estimate $2\left<\sigma_+\right> = \left<X+i\;Y\right>$ from which we can approximate $\frac1{d}\text{Tr}(U)$.}
    \label{fig:trU-circuit}
\end{figure}

In principle, the need for the use of random samplings can be circumvented if each stochastic vector has a uniform overlap with the eigenstates of $H$. We here propose and implement a protocol that employs the time evolution on quantum computers with random vectors drawn from such a distribution, surpassing the classical limitations of the stochastic implementation. We argue that this protocol is amenable to implementation on near-term quantum computers. 

This protocol entails constructing a maximally mixed state $\rho_{\rm MMS}$ (shown in Fig.~\ref{fig:trU-circuit}a) as the input to a computational register, in
contrast to an initial pure state (which may be prepared to have \textit{some} overlap on the relevant subspace), for usual time evolution on a quantum computer controlled by the {\it pointer} qubit, depicted in Fig.~\ref{fig:trU-circuit}b. While the protocol can provide input for many quantum algorithms, including QPE~\cite{Abrams1997, Lloyd1996}, here we apply time evolution via the Hadamard test to calculate eigenvalues.  We also note that this technique can be easily extended to correlation functions in general~\cite{Somma2019,Somma2002}, because hardware requirements (e.g., error rate, circuit depth, coherence time) are within the capabilities of current noisy quantum computers for time-series experiments on small enough computational registers~\cite{Preskill2018}. From here, the usual $X$- and $Y$-basis measurements as a function of time are used to construct the time series (see \hyperref[app:hadamard-test]{Appendix})
\begin{align}
    \langle X + iY \rangle(t) = 2\langle \sigma^+ \rangle (t) = \frac1{d}{\rm Tr}(U(t)),
\end{align}
which determines the spectrum through a fourier transformation. In our implementation we create the maximally mixed state by entangling each qubit in the computational  register with a ``{\it garbage}" qubit which remains unmeasured (and presumably coherent) throughout the algorithm execution.

\section{Eigenvalue Spectrum Estimation}\label{section:spectrum}
As a concrete example to contrast the stochastic sampling and quantum evolution techniques, we consider
the general one-dimensional spin-1/2 Heisenberg Hamiltonian with a $\hat{\mathbf{z}}$ direction external field coupling, which has the form
\begin{equation}
\hat{H} = -J\sum_{\langle ij \rangle} \left(X_iX_j + Y_iY_j+ Z_iZ_j \right) - B\sum_i Z_i, 
\end{equation}
where $X_i, Y_i$, and $Z_i$ are Pauli operators at site $i$, $\langle ij \rangle$ sums nearest-neighbor (NN) sites, with a coupling strength $J$. At each site, electrons can take spin up $\ket{\uparrow} (\equiv \ket{0})$ or spin down $\ket{\downarrow} (\equiv \ket{1})$ configuration in the Pauli-$Z$ basis.  Physically, the Pauli operators $X_i$ and $Y_i$ generate spin flips between NN sites, leading to the exchange interaction between spins. The $Z_i$ operator, on the other hand, favors parallel alignment between NN sites, resulting in a ferromagnetic coupling.

In this work, we asses the performance of the quantum algorithm running on a  noisy quantum hardware in comparison to stochastic methods, by implementing a 2-site Heisenberg model. In this case, the Hamiltonian can be represented in a matrix form
\begin{equation}
    H=-\begin{bmatrix}
        J+2B &  0 & 0 & 0\\
        0&-J& 2J & 0\\
        0 & 2J &-J &0\\
        0 & 0 & 0& J-2B
    \end{bmatrix},
\end{equation}
in the Pauli-$Z$ basis $\{\ket{\downarrow\downarrow},\ket{\uparrow\downarrow},\ket{\downarrow\uparrow},\ket{\uparrow\uparrow}\}$. Its eigenvalues eigenvalues are, $-J \pm 2B $ and $J\pm 2J$, where degeneracy occurs when the external field $B=\pm 2J$.

\subsection{Quantum Hardware Approach}\label{subsec:quantum-approach}

\subsubsection{Preparing a maximally mixed state}
\begin{figure}
    \centering
    \includegraphics[width=0.4\textwidth]{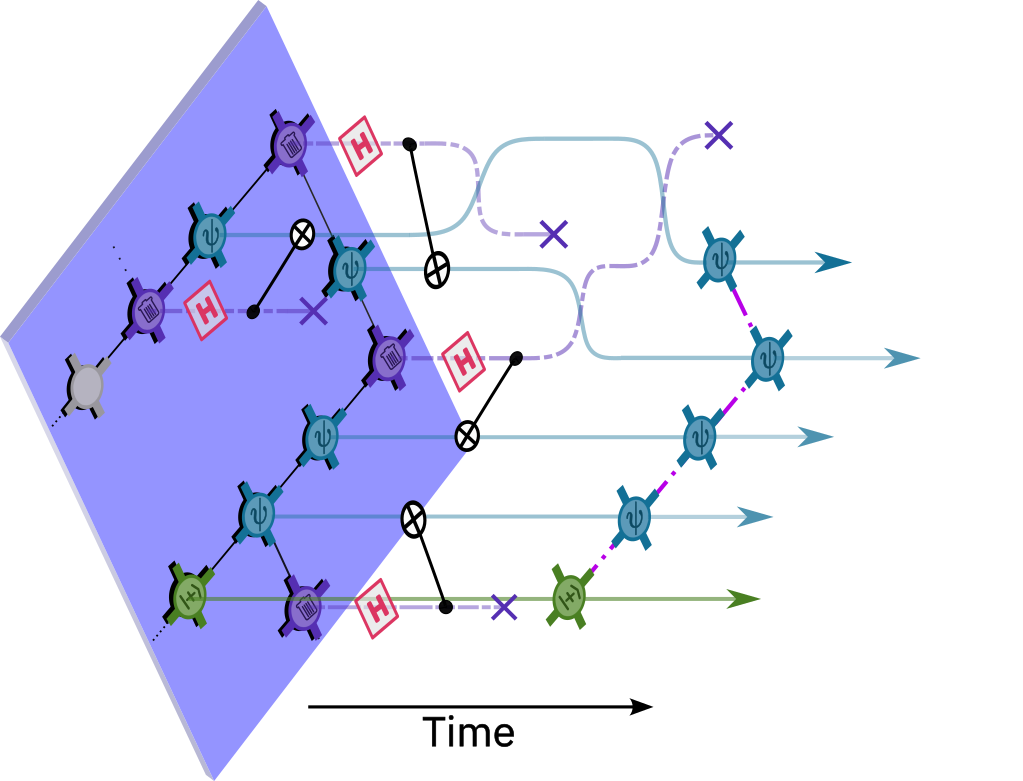}
     \caption{ {\bf Schematic depiction of preparing a maximally mixed state of 4 qubits on the heavy-hexagonal topology of IBM Quantum's superconducting processors}.  The purple qubits and dashed $X$ lines represent the garbage qubits while the teal qubits and arrows represent the computation qubits.  The pointer qubit is represented in green with $\ket{+}$. Due to the connectivity, three \texttt{SWAP} gates (represented as crossed wires in the figure) are required in this state preparation protocol.  } 
     \label{fig:MMS-figure}
\end{figure}

A maximally mixed qubit register is experimentally prepared by preparing it, together  with a garbage qubit register, in one of the Bell states, for example, using the circuit depicted in Fig.~\ref{fig:trU-circuit}a.  Ideally, the garbage qubit should be left unmeasured and protected against relaxation processes during the computation. To understand why, consider the state $\ket{\phi^{+}}=(\ket{0}_c\ket{0}_g+\ket{1}_c\ket{1}_g)/\sqrt2$. If one of the qubits is subjected to a measurement, then the other qubit is projected onto the same state. In such a case, instead of evolving the maximally mixed state, only a single computational basis state is evolved. Note, that it is enough to protect the qubit against any mechanism that corresponds to a measurement process. On the other hand if the garbage qubit goes through a trace preserving quantum channel $\cal{E}(\cdot)$, then the reduced state of the computation qubit is
\begin{equation}
\rho_c = \frac1{2}\sum_{i,j=0}^1{\rm Tr}[{\cal E}(\ket{i}_g\!\bra{j})] \ket{i}_c\!\bra{j}  = \frac1{2}\mathbf{I}.    
\end{equation}
Therefore, the preparation of the maximally mixed state is robust against depolarizing channel or other trace-preserving channels acting  on the garbage qubit.

Another consideration in preparing the maximally mixed state is the limited connectivity between qubits in planar architectures. As controlled-NOT \texttt{CX} gates are the
major source of error in superconducting qubit architectures, with error rates in the range of 0.5-1\%, it is important to minimize the number of \texttt{SWAP} gates (consisting of up to three \texttt{CX} gates). A diagram depicting how minimizing the number of \texttt{CX} gates may be achieved for a one-dimensional line of qubits on IBM's heavy-hexagonal topology~\cite{Chamberland2019} is displayed in Fig.~\ref{fig:MMS-figure}.

\begin{figure*}
\begin{flushleft}
    {\textbf{(a)}}~
    \Qcircuit @C=1em @R=.5em {
        & \ctrl{1} & \qw \\
        & \multigate{1}{R_{zz}(\theta)} & \qw  \\
        & \ghost{R_{zz}(\theta)} & \qw
        } \quad
    {\rm (i)}~
        \Qcircuit @C=1em @R=.7em {
        & \ctrl{2} & \qw & \ctrl{2} & \qw & \ctrl{2} & \ctrl{2} \\
        \lstick{ =\ \ } &  \ctrl{1} & \qw & \qw & \qw & \qw & \ctrl{1} \\
        & \targ & \gate{R_z\left(\frac{\theta}{2}\right)} & \targ & \gate{R_z\left(\frac{-\theta}{2}\right)} & \targ & \targ
        } \quad \vline \quad
    {\textbf{(d)}}~
    \Qcircuit @C=0.7em @R=2.0em {
        & \ctrl{2} &\qw \\
        & \ctrl{1} &\qw  \\
        & \targ & \qw
    } \qquad
    \Qcircuit @C=0.5em @R=0.9em {
        & \qw & \qw & \qw  &\ctrl{2} & \qw & \qw & \qw & \ctrl{2} & \ctrl{1} & \gate{T} & \ctrl{1}\\
        \lstick{=}& \qw & \ctrl{1} &\qw & \qw & \qw & \ctrl{1} &\gate{T} &\qw & \targ &\gate{T^\dagger} & \targ\\
        & \gate{H} & \targ  &\gate{T^\dagger} & \targ & \gate{T} & \targ & \gate{T^\dagger} & \targ & \gate{T} & \gate{H} & \qw
    } \quad
\end{flushleft}
\begin{tikzpicture}
\draw[gray, opacity=0.7, line width=0.3mm] (-15,1) -- (2,1);
\end{tikzpicture}

\begin{flushleft}
    {\textbf{(b)}}~
    \Qcircuit @C=1em @R=.5em {
        & \ctrl{1} & \qw \\
        & \multigate{1}{R_{zz}(\theta)} & \qw \\
        & \ghost{R_{zz}(\theta)} & \qw
        } \quad
    {\rm (i)}~
    \Qcircuit @C=1em @R=.7em {
        & \qw & \ctrl{1} & \qw & \ctrl{1} & \qw \\
       \lstick{=\ \ } & \multigate{1}{R_{zz}(\theta/2)} & \targ & \multigate{1}{R_{zz}(-\theta/2)} & \targ & \qw & \\
        & \ghost{R_{zz}(\theta/2)} & \qw & \ghost{R_{zz}(-\theta/2)} & \qw & \qw
        } \quad
        {\rm (ii)}~
        \Qcircuit @C=1em @R=.7em {
        & \qw & \qw & \qw & \ctrl{1} & \qw & \qw & \qw & \ctrl{1} \\
       \lstick{ =\ \ \ } & \ctrl{1} & \qw & \ctrl{1} & \targ & \ctrl{1} & \qw & \ctrl{1} & \targ \\
        & \targ & \gate{R_z\left(\frac{\theta}{2}\right)} & \targ & \qw & \targ & \gate{R_z\left(\frac{-\theta}{2}\right)} & \targ & \qw
        \gategroup{2}{2}{3}{4}{1.4em}{--}
        \gategroup{2}{6}{3}{8}{1.4em}{--}
        } \quad 
\end{flushleft}

\begin{tikzpicture}
\draw[gray, opacity=0.7, line width=0.3mm] (-15,1) -- (2,1);
\end{tikzpicture}

\begin{flushleft}
{\textbf{(c)}}~
    \Qcircuit @C=1em @R=.7em {
    & \ctrl{1} & \qw \\
    & \multigate{1}{R_{zz}(\theta)} & \qw  \\
    & \ghost{R_{zz}(\theta)} & \qw
    } \quad 
    {\rm (i)}~
    \Qcircuit @C=.5em @R=.7em {
    & \qw & \ctrl{1} & \qw & \qw \\
    \lstick{=\ \ }& \targ & \gate{R_z(\theta)} & \targ & \qw \\
    & \ctrl{-1} & \qw & \ctrl{-1} & \qw 
    } \quad \vspace{0.6em}
    {\rm (ii)}~
    \Qcircuit @C=0.4em @R=.7em {
    & \lstick{} & \qw & \qw & \multigate{1}{R_{zz}(-\theta/2)} & \qw & \qw \\
    &\lstick{=\ \ } & \targ & \gate{R_z(\theta/2)} & \ghost{R_{zz}(-\theta/2)} & \targ & \qw \\
    & \lstick{} & \ctrl{-1} \qw & \qw & \qw & \ctrl{-1} & \qw 
    } \quad
    {\rm (iii)}~
    \Qcircuit @C=0.8em @R=.7em {
    & \qw & \qw & \ctrl{1} & \qw & \ctrl{1} & \qw \\
    \lstick{=\ \ } & \targ & \gate{R_z(\theta/2)} & \targ & \gate{R_z(-\theta/2)} & \targ & \targ \\
    & \ctrl{-1} & \qw & \qw & \qw & \qw & \ctrl{-1} \gategroup{1}{4}{2}{6}{1.2em}{--} 
    } \quad
\end{flushleft}
\caption{{\bf Various decompositions of a ctrl-$R_{zz}(\theta)$ gate} (note that any other two-qubit Pauli rotation can be decomposed to $R_{zz}(\theta)$ by placing the appropriate single-qubit gates on either side of the operation).  {\bf(a)} The default decomposition using Qiskit's standard equivalence library in a\textit{(i)} contains two Toffoli gates which decompose to six \texttt{CX} gates {\bf(d)} is very inefficient.  Alternatively in {\bf(b)}, the controlled $R_{zz}(\theta)$ gate can be written using two $R_{zz}(\theta)$ gates as shown in b\textit{(i)}.  In {\bf(c)} we show the decomposition to a single ctrl-$R_z(\theta)$ gate between two \texttt{CX} gates c\textit{(i)}.  The single ctrl-$R_z(\theta)$ gate is then expressed by a $Z$-rotation for half of the total angle $\theta$ followed by an $R_{zz}(-\theta/2)$ rotation in c\textit{(ii)} which realizes a circuit with a minimal amount of two-qubit gates.  In both {\bf(b)} and {\bf(c)}, the $R_{zz}(\theta)$ gates can be expressed as a pair of \texttt{CX} gates between a single $R_z(\theta)$ (which are realized virtually via frame changes ~\cite{McKay2017} and therefore noiseless) as shown in boxes in {\bf (b)}\textit{(ii)} and {\bf (c)}\textit{(iii)}. Here $R_{zz}(\theta)$ rotations can be constructed by pulse scaling to reduce the effective gate time~\cite{Stenger2021} }
    \label{fig:zzs}
\end{figure*}
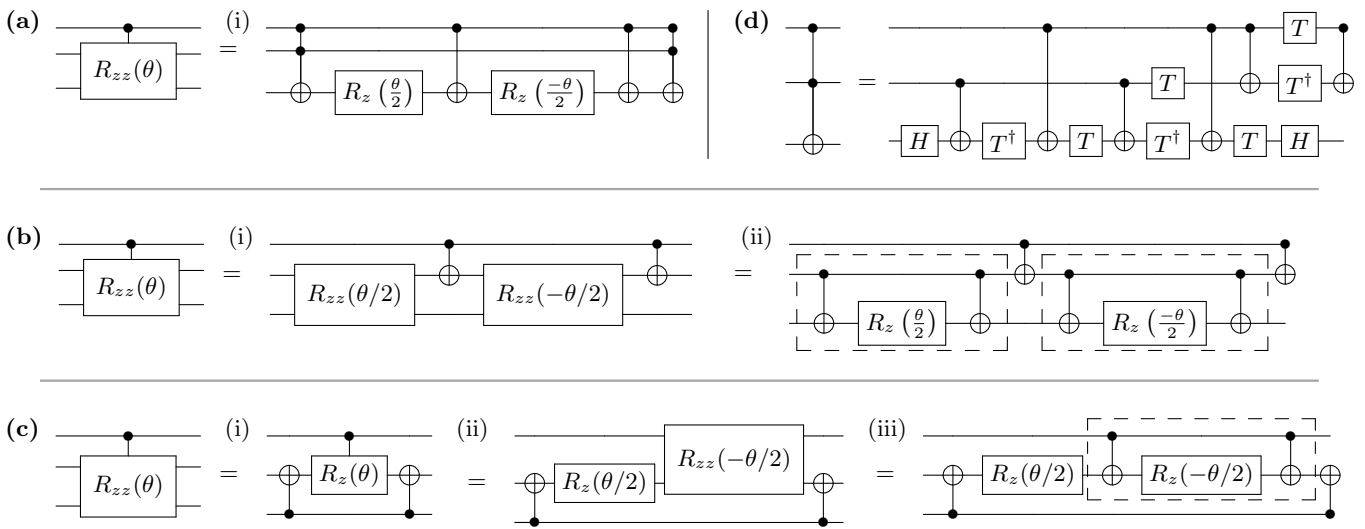

\subsubsection{Implementation of controlled time-evolution}
Our protocol requires an implementation of a controlled time evolution circuit. However, the decomposition of an arbitrary controlled unitary circuit using Qiskit's default equivalence library synthesizes gate sequences~\cite{shende-2006} that cannot be realistically executed (see Fig.~\ref{fig:zzs}a(i)) on noisy near-term quantum hardware, even for the modest 2-qubit Hamiltonian we are considering here. The limited connectivity between qubits potentially requires \texttt{SWAP} gates, which additionally provides a challenge to minimizing the number of \texttt{CX} gates. 

To overcome this, we first realize that, up to single-qubit rotations,  any controlled-Pauli rotation is locally equivalent to 
\begin{equation}
cP_{R(\theta)}  \sim  \ket{0}\bra{0} + e^{-i\frac{\theta}{2}\bigotimes_{j} P^i_{j}} \ket{1}\bra{1}
\end{equation}
where $P^i \in \{I, Z\}$ consisting of a tensor product of identity and Pauli-$Z$ operators, often referred to in the {\it Pauli string} notation where position in the string indicates which qubit is operated on. These Pauli rotations can be realized as a single-qubit $R_z(\theta)$ on a single {\it choice} of qubit with a $Z$ term in the Pauli string, and \texttt{CX}s targeting this qubit. It is controlled on every other qubit with a $Z$ in the Pauli string, both before and after the $R_z(\theta)$. Therefore, when constructing the controlled-rotation, one is free to place the $R_z(\theta)$ on the qubit with the least distance to the pointer qubit within the connectivity graph of the quantum computer, as this reduces the number of required \texttt{SWAP}s (and hence \texttt{CX}s). 

Next, we implement the controlled $Z$-rotation via an {\it uncontrolled} rotation by half the angle $R_z(\theta/2)$ followed by an $R_{zz}(-\theta/2)$ gate. The single-qubit $Z$-rotation is implemented without error by a frame change~\cite{McKay2017}. The two-qubit $ZZ$-rotation can be implemented with scaled echoed cross resonance gates~\cite{Stenger2021}. This technique allows us to reduce the number of \texttt{CX} gates by 75\% (from roughly 100 to about 25), for the specific Hamiltonian we are considering (Table~\ref{table:cx-count}).  Performing the uncontrolled evolution first is important as it allows the preparation of the superposition on the pointer qubit (by Hadamard gate) as late as possible, preventing unnecessary decoherence.

\begin{table}
\begin{center}
\begin{tabular} {c || c | c | c}
\hline \hline
    Synthesis & CX Count & CX Count(mapped) & ECR Count \\
    \hline \hline
    Fig. \ref{fig:zzs}a & 50 & 67 & 104 \\
    \hline
    Fig. \ref{fig:zzs}b & 22 & 23 & 18 \\
    \hline
    Fig. \ref{fig:zzs}c & 16 & 16 & 15 \\
    \hline
\end{tabular}
\end{center}
\caption{CX count for each circuit synthesis technique shown in Figure~\ref{fig:zzs} which constructs the controlled time evolution of the Heisenberg dimer.  The values shown are counted before and after mapping to the qubit connectivity of the device. The number of echoed cross-resonance (ECR) pulses (both scaled and unscaled) after mapping is also listed for these synthesis techniques.}
\label{table:cx-count}
\end{table}

The careful decomposition of the controlled time-evolved Heisenberg Hamiltonian is highlighted in Fig.~\ref{fig:zzs} by considering just the $ZZ$ term.  Fig.~\ref{fig:zzs}a shows a decomposition first using Qiskit's standard equivalence library for the $R_{zz}(\theta)$ gate followed by the addition of controls on each individual gate. Fig.~\ref{fig:zzs}b shows a decomposition by controlling the direction of rotation of a $R_{zzz}(\theta/2)$ gate (consisting of an $R_{zz}(-\theta/2)$ sandwiched between 2 \texttt{CX}s) followed by an action of the equivalence library on the $R_{zz}$ gates in the dashed boxes. The most hardware-efficient decomposition appears in Fig.~\ref{fig:zzs}c where the control is placed on the single-qubit $R_z(\theta)$ necessary to achieve the Pauli rotation, which is further decomposed into an error-free $R_z(\theta/2)$ implemented virtually~\cite{McKay2017} and an $R_{zz}(-\theta/2)$ gate that produced by scaling the native cross resonance interaction~\cite{Stenger2021}.

\section{Results}\label{section:results}

\begin{figure}
    \centering
    \begin{flushleft}
(a)
    \end{flushleft}
     \includegraphics[width=0.45\textwidth]{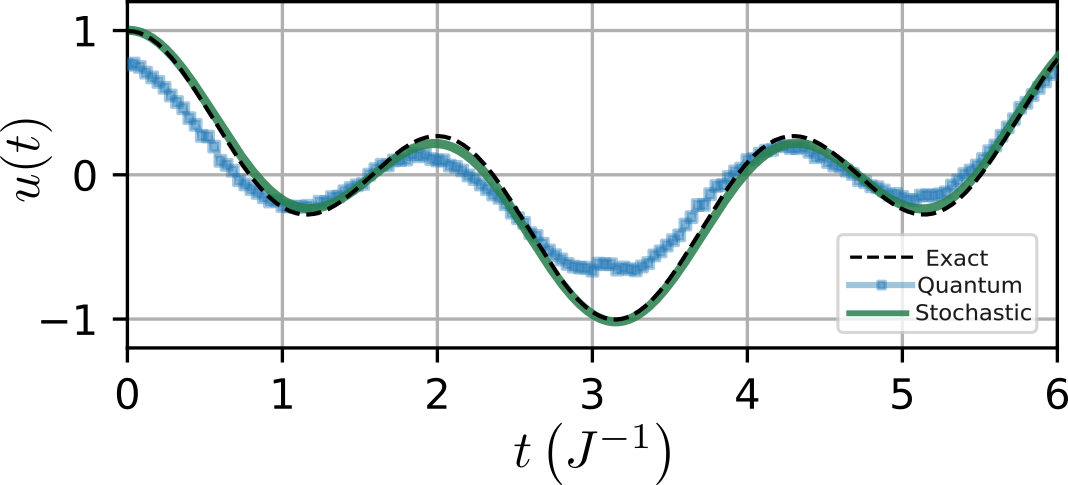}
\begin{flushleft}
(b)
    \end{flushleft}
    \includegraphics[width=0.45\textwidth]{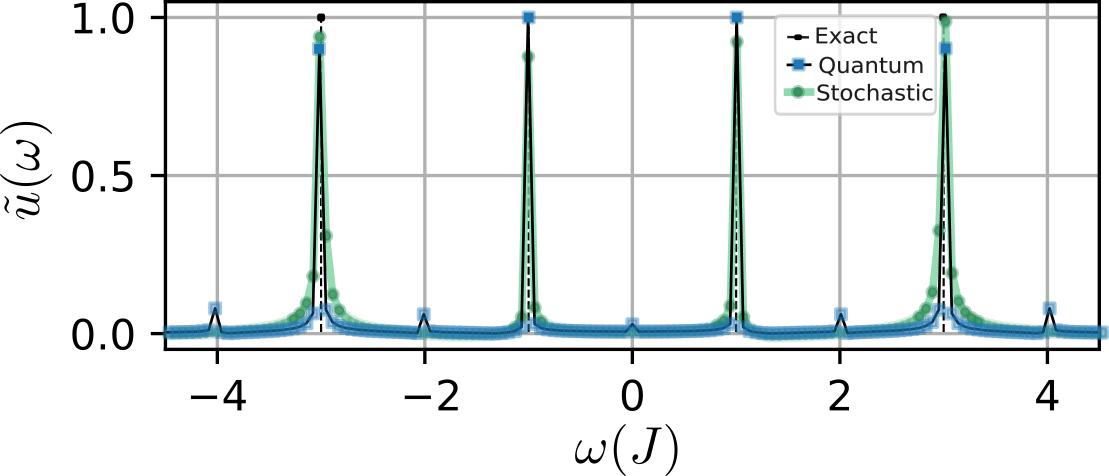}
     \caption {{\bf Spectral comparison between a classical stochastic method and controlled quantum propagation on a 2-site Heisenberg model} (at $J/B \equiv 1$). (a) The time evolution amplitudes: the dash-line in black represents the \textit{exact} amplitude, e.g. $u(t)$;  The lines in blue are expectation values of the Pauli-Y operator ($\braket{Y}$), acquired from the ancilla qubit; The line in green labels the amplitude generated by the stochastic technique, with sample size $S=100$. The unit time step for the classical evolution is $dt=10^{-4}~J^{-1}$, $T=6~J^{-1}$, and for easy comparison, we project the \textit{ibm\_lagos} data onto an identical time and frequency grid used in the classical algorithm via a quadratic interpolation. (b) The corresponding normalized power spectrum generated by the discrete Fourier transform (DFT), with the same color labeling. Both spectra generated are in close agreement with the exact spectrum (black vertical dash-line).  The harmonic peaks at even integer frequencies are artifacts generated by the frequency interpolation scheme via the DFT.} 
     \label{fig:compare-classical-and-quantum-spectrum}
\end{figure}

To assess our eigenvalue estimation approach, we compare the spectral decomposition of a Heisenberg dimer with an energy ratio $J/B=1$ between the classical stochastic method and the quantum hardware approach.  We obtain this first by stochastically sampling the 2-site Hilbert space with $S=100$ samples and evolve each sampled state to time $T\approx2\pi J^{-1}$ with time steps $dt=10^{-4} J^{-1}$ following the common implementation, as detailed, e.g., in Ref.~\cite{StochasticReview}.   The resulting average expectation value of $\langle \psi_{\rm stoc}^{(k)}(t)\rangle$ and eigenvalue spectrum $\tilde\psi_{K}(\omega)$ are then normalized and plotted in green in Figs.~\ref{fig:compare-classical-and-quantum-spectrum}a and \ref{fig:compare-classical-and-quantum-spectrum}b respectively -- along with the exact evolution.

\begin{figure}
    \centering
    \includegraphics[width=0.45\textwidth]{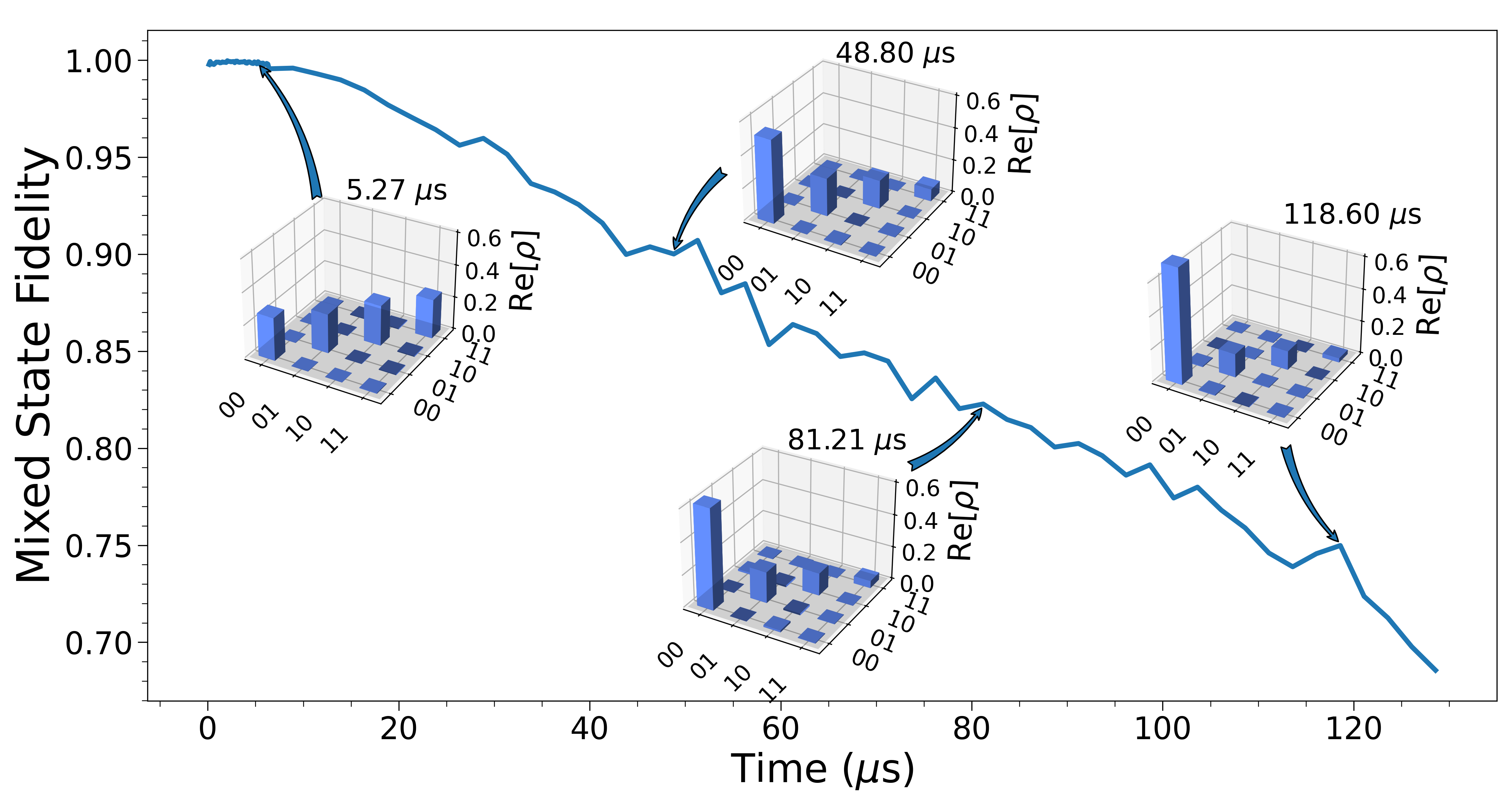}
    \caption{{\bf Lifetime of maximally mixed state} (a) The fidelity between the state of the computational qubits with the maximally mixed state as a function of time.  Two garbage qubits are prepared in a maximally entangled state with two computational qubits,  and left idle for time $t$ before tomographic measurement. The probability of obtaining the $|00\rangle$ state increases with time due to qubit relaxation.}
    \label{fig:mms-lifetime}
\end{figure}

For the quantum approach, the time evolution was simulated using the 7-qubit \textit{ibm\_lagos} device and the controlled-unitary $U_c(t)$ was generated using the approach described in this work.  We first prepared two Bell pairs then executed a controlled evolution of two qubits (one from each pair) under a 2-site Heisenberg Hamiltonian.  The same energy ratio of $J/B=1$ was chosen and the system was evolved for time $T\approx 2\pi~J^{-1}$ and $dt\approx0.04~J^{-1}$ under the first order Trotter-Suzuki approximation with a single Trotter step.  For each time step, a number of shots $N = 8192$ of the observable $\langle X + iY \rangle$ was performed.  Experimentally, the system suffers from a non-trace-preserving relaxation channel which will destroy the maximally-mixed state between the computation and garbage qubits.  To ensure our circuits are much shorter than this decay, in Fig.~\ref{fig:mms-lifetime} we show the effective lifetime of the two-qubit maximally mixed state by preparation and tomographic measurement of the computational register after idle time $t$.  Given that our experiments execute within 9.5$\mu s$, the fidelity of the maximally mixed state is high enough such that our analysis will return accurate eigenvalue spectra.

If perfect single-shot measurement is assumed, the standard deviation of the expectation of an observable is $\Delta \langle \mathcal{O} \rangle = 2/\sqrt{N}$, where we require 2: $\langle X \rangle$ and $\langle Y \rangle$, giving standard deviation $1/\sqrt{N}$ overall. Given a uniform sampling over the Hilbert space $d=2^N$ we can select windows of time evolution to lessen the degradation to the signal from measuring multiple eigenvalues.  Additionally, to mitigate hardware noise, we execute all $R_{zz}(\theta)$ gates via cross-resonance pulse scaling~\cite{Stenger2021}.  By doing so, the error rate of these two-qubit Pauli rotations is linearly scaled according to the angle $\theta$.   The data was then projected onto an identical time grid to the stochastic approach via a quadratic interpolation and processed by a discrete Fourier transform.  The corresponding normalized eigenvalue spectrum is shown in blue in Fig.~\ref{fig:compare-classical-and-quantum-spectrum}b.

The precision of unitary evolution techniques is limited by the total time evolved, which is ultimately limited by the number of gates that can be completed within the qubit coherence times. With this limitation, however, results agree well compared to classical numerics for $T=6$. Further improvements are expected as qubit coherence times increase.

\begin{figure}
    \centering
    \includegraphics[width=0.51 \textwidth]{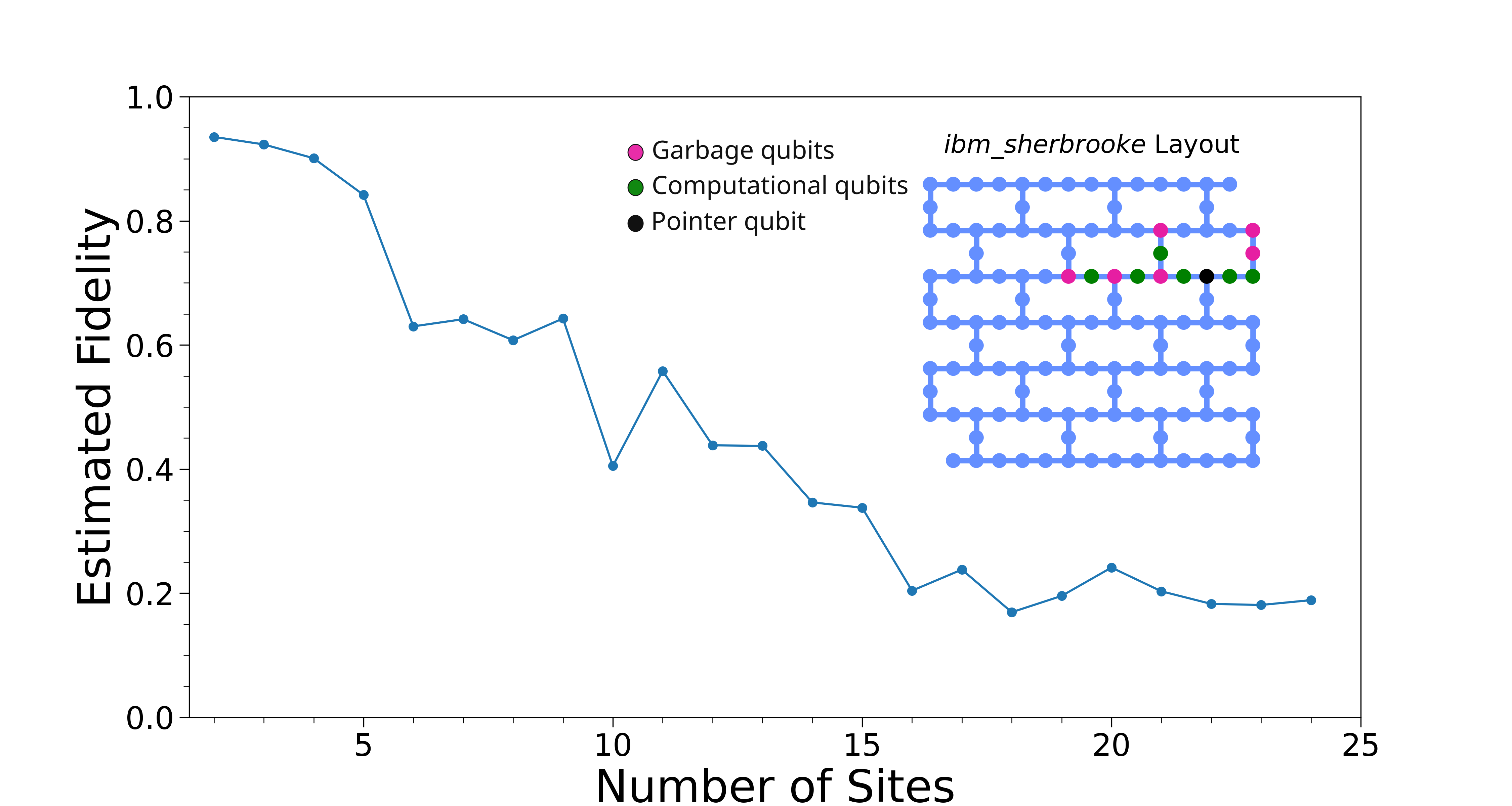}
    \caption{ {\bf Estimated fidelity of time evolution as a function of chain length.}  For each chain we prepare a maximally mixed stated and the controlled time evolution protocol then estimate the fidelity by taking a produce of the error rates obtained from the device calibration.  Inset shows a connectivity graph of the \textit{ibm\_sherbrooke} device with a sample layout.  The computation qubits are highlighted in green while the ``garbage" qubits used to prepare the MMS are shown in pink.  The phase qubit used to measure $\langle X + iY \rangle$ is shown in black.}
    \label{fig:est_fid_chain}
\end{figure}

As further analysis, we estimate the fidelity of this controlled time evolution for large chain lengths in Fig.~\ref{fig:est_fid_chain}.  To obtain these estimates, we prepare the eigenvalue spectrum estimation protocol described in Section~\ref{subsec:quantum-approach} for a given chain length of $n$.  The routing of the associated circuit is then optimized to \textit{ibm}\_\textit{sherbrooke}, a 127-qubit device configured to a heavy-hex connectivity graph.  Finally, the calibration information of the device is gathered to implement the weight-2 Pauli rotations (as shown in Fig.~\ref{fig:zzs}).  An estimated fidelity was then calculated using the prepared circuit -- evolving the system to time $t\approx2\pi~J^{-1}$ with the same energy ratio and time evolution approximation -- by taking a product of the error rates obtained from the device calibration for all of the single and two-qubit gates.  We also take into account the linear scaling of the error rates for the cross-resonance pulses for this estimate~\cite{Stenger2021}.  The results here demonstrate an estimated fidelity less than $0.5$ once the chain has grown to $10$ sites.  An inset shows a sample layout of a small,  6-site chain on the \textit{ibm\_sherbrooke} device with the system qubits highlighted in green and the ``garbage" qubits highlighted in pink.  Additionally, we note that because the transpilation procedure utilizes a probabilistic heuristic to insert needed \texttt{SWAP} gates~\cite{sabre-swap}, the fidelity estimate possesses some variance in what should be a smooth function. 

\section{Perspective and Conclusions}\label{section:conclusions}

This work explores the practical advantages of quantum realization of randomized sampling for spectral estimation, which (similar to the classical counterpart) enables exploiting the information redundancy. In essence, the classical approach requires multiple re-samplings, while the quantum realization allows constructing ``ideal’’ (uniformly distributed) random states even on near-term devices and yields an unbiased estimation of correlation functions.

The quantum resources required for this sampling technique scale favorable with system size. The time step sets the cutoff frequency and thereby largest eigenvalue that can be observed, whereas the total time evolution sets both the minimum eigenvalue that can be observed as well as the minimum resolution between eigenvalues. The total time evolution is currently limited by qubit decoherence, as circuit depth increases with larger number of Trotter steps. Neglecting measurement error, which maybe be mitigated~\cite{Nation2021}, the signal will be limited due to shot noise to a standard deviation of $1/\sqrt{N}$ where $N$ is the number of identical experiments performed for each the $X$ and $Y$ expectation values. 
For example, here $N=8192$ experiments were performed twice for each time step at a typical rate of 4~kHz. 
This signal is further suppressed due to spectral weight transfer among all the eigenstates contributing to the time evolution. Since we are typically interested only in $\tilde d$ fraction of the Hilbert space explored, the signal suppression and shot noise contribute overall as $\sim \tilde{d}/\sqrt{N}$ to the overall error. Further improvement is possible via denoising through projecting the measured data onto the ``nearest'' function satisfying the physical constraints of the time correlators, as has been recently demonstrated for the equilibrium and steady states~\cite{kemper2023positive}.

Besides the hardware advances necessary to achieve longer time evolution (and hence better frequency resolution), the near-term strategy may alternatively employ numerical postprocessing on classical devices. In particular, techniques based on principle mode analysis and dimension reduction technique through low-rank approximations (such as dynamical mode decomposition \cite{DMD0,schmid2011applications,TuRowley,DMDVV, DMDdiag, mejía2023stochastic}). Alternative strategies may employ, for instance, Recursive Neural Networks, in which the general functional forms of the time evolution are trained on solvable model.\cite{bassi2023learning}. Such a combination of multiple techniques is likely optimal for practical applications in the near future and circumvents the need for a huge technological leap.

In this work, we have practically demonstrated an approach for estimating the spectrum of a quantum many-body Hamlitonian.  We utilize a novel technique to efficiently construct a controlled time evolution operator $U_c(t)$. Our results show agreement with classical stochastic methods of obtaining eigenvalue spectra and demonstrate the ability to accurately reproduce eigenvalue spectra for a small Heisenberg system.  With the introduction of more sophisticated error mitigation techniques~\cite{Temme2017, Sung2022, Kim2023, VandenBerg2023, Majumdar2023} and a direct measurement protocol for the observable measurement~\cite{Mitarai2019}, quantum hardware may be able to outperform classical stochastic methods of obtaining eigenvalues of larger systems -- allowing for the accurate characterization of condensed matter models currently inaccessible using numerical, stochastic, or other approximating methods.

\section{Acknowledgment}
This research on random sampling (VV and ZW) was supported NSF CAREER award DMR 1945098. IH and AK would like to acknoledge support from NSF award \#2210374. The authors thank Jeffrey Cohn for a careful reading of the manuscript and acknowledge the use of IBM Quantum Services for this work. The collaboration was enabled through the collaboration with Quantum Foundry at UCSB (NSF award DMR-1906325). In addition, this material is based upon work supported
by the Defense Advanced Research Projects Agency (DARPA) under Contract No. HR001122C0063. 

\phantomsection
\section*{Appendix - Hadamard test}~\label{app:hadamard-test}

Very generally, the Hadamard test concerns with the application of a controlled-unitary operation, that is the application of a unitary transformation $U$ on register qubits, conditioned that an ancilla qubit is in a $\ket{1}$ state. By applying a controlled unitary $U_c$ on a state $\rho$ when the control qubit is prepared in the $\ket{+}=(\ket{0}+\ket{1})/\sqrt2$ we obtain:
\begin{widetext}
$$U_c(\ket{+}\bra{+}\otimes \rho) U_c^\dagger=
\frac1{2}\Big[\ket{0}\bra{0}\otimes \rho+
    (\ket{0}\bra{1}\otimes (\rho U^\dagger)+
    \ket{1}\bra{0}\otimes (U\rho)+
    \ket{1}\bra{1}\otimes (U\rho U^\dagger)\Big]$$
For the special case where $\rho$ is the completely mixed state we have
$$U_c(\ket{+}\bra{+}\otimes \rho) U_c^\dagger=
\frac1{2d}\Big[\ket{0}\bra{0}\otimes\mathbf{I}+
    (\ket{0}\bra{1}\otimes U^\dagger+
    \ket{1}\bra{0}\otimes U+
    \ket{1}\bra{1}\otimes\mathbf{I})\Big].$$
    Therefore, the reduced state of the qubit after the application of the controlled unitary operation becomes
\begin{equation*}
\rho_{\rm qubit}=
\frac1{2}\mathbf{I}+\frac1{2d}\Big[\ket{0}\bra{1} \text{Tr}(U^\dagger)+
    \ket{1}\bra{0}\text{Tr}(U)\Big]
\end{equation*}
Hence measuring $X+iY=2\ket{1}\bra{0}$ corresponds to measuring  $\text{Tr}(U)/d$.
\end{widetext}

\bibliography{stochastic-somma-refs}

\begin{thebibliography}{43}%
\makeatletter
\providecommand \@ifxundefined [1]{%
 \@ifx{#1\undefined}
}%
\providecommand \@ifnum [1]{%
 \ifnum #1\expandafter \@firstoftwo
 \else \expandafter \@secondoftwo
 \fi
}%
\providecommand \@ifx [1]{%
 \ifx #1\expandafter \@firstoftwo
 \else \expandafter \@secondoftwo
 \fi
}%
\providecommand \natexlab [1]{#1}%
\providecommand \enquote  [1]{``#1''}%
\providecommand \bibnamefont  [1]{#1}%
\providecommand \bibfnamefont [1]{#1}%
\providecommand \citenamefont [1]{#1}%
\providecommand \href@noop [0]{\@secondoftwo}%
\providecommand \href [0]{\begingroup \@sanitize@url \@href}%
\providecommand \@href[1]{\@@startlink{#1}\@@href}%
\providecommand \@@href[1]{\endgroup#1\@@endlink}%
\providecommand \@sanitize@url [0]{\catcode `\\12\catcode `\$12\catcode
  `\&12\catcode `\#12\catcode `\^12\catcode `\_12\catcode `\%12\relax}%
\providecommand \@@startlink[1]{}%
\providecommand \@@endlink[0]{}%
\providecommand \url  [0]{\begingroup\@sanitize@url \@url }%
\providecommand \@url [1]{\endgroup\@href {#1}{\urlprefix }}%
\providecommand \urlprefix  [0]{URL }%
\providecommand \Eprint [0]{\href }%
\providecommand \doibase [0]{https://doi.org/}%
\providecommand \selectlanguage [0]{\@gobble}%
\providecommand \bibinfo  [0]{\@secondoftwo}%
\providecommand \bibfield  [0]{\@secondoftwo}%
\providecommand \translation [1]{[#1]}%
\providecommand \BibitemOpen [0]{}%
\providecommand \bibitemStop [0]{}%
\providecommand \bibitemNoStop [0]{.\EOS\space}%
\providecommand \EOS [0]{\spacefactor3000\relax}%
\providecommand \BibitemShut  [1]{\csname bibitem#1\endcsname}%
\let\auto@bib@innerbib\@empty
\bibitem [{\citenamefont {Martin}\ \emph {et~al.}(2016)\citenamefont {Martin},
  \citenamefont {Reining},\ and\ \citenamefont {Ceperley}}]{martin_2016}%
  \BibitemOpen
  \bibfield  {author} {\bibinfo {author} {\bibfnamefont {R.~M.}\ \bibnamefont
  {Martin}}, \bibinfo {author} {\bibfnamefont {L.}~\bibnamefont {Reining}},\
  and\ \bibinfo {author} {\bibfnamefont {D.~M.}\ \bibnamefont {Ceperley}},\
  }\href {https://doi.org/10.1017/CBO9781139050807} {\emph {\bibinfo {title}
  {Interacting Electrons: Theory and Computational Approaches}}}\ (\bibinfo
  {publisher} {Cambridge University Press},\ \bibinfo {year}
  {2016})\BibitemShut {NoStop}%
\bibitem [{\citenamefont {Baer}\ \emph {et~al.}(2022)\citenamefont {Baer},
  \citenamefont {Neuhauser},\ and\ \citenamefont {Rabani}}]{StochasticReview}%
  \BibitemOpen
  \bibfield  {author} {\bibinfo {author} {\bibfnamefont {R.}~\bibnamefont
  {Baer}}, \bibinfo {author} {\bibfnamefont {D.}~\bibnamefont {Neuhauser}},\
  and\ \bibinfo {author} {\bibfnamefont {E.}~\bibnamefont {Rabani}},\
  }\bibfield  {title} {\bibinfo {title} {Stochastic vector techniques in
  ground-state electronic structure},\ }\href
  {https://doi.org/10.1146/annurev-physchem-090519-045916} {\bibfield
  {journal} {\bibinfo  {journal} {Annual Review of Physical Chemistry}\
  }\textbf {\bibinfo {volume} {73}},\ \bibinfo {pages} {255} (\bibinfo {year}
  {2022})},\ \bibinfo {note} {pMID: 35081326},\ \Eprint
  {https://arxiv.org/abs/https://doi.org/10.1146/annurev-physchem-090519-045916}
  {https://doi.org/10.1146/annurev-physchem-090519-045916} \BibitemShut
  {NoStop}%
\bibitem [{\citenamefont {Neuhauser}\ \emph {et~al.}(2014)\citenamefont
  {Neuhauser}, \citenamefont {Gao}, \citenamefont {Arntsen}, \citenamefont
  {Karshenas}, \citenamefont {Rabani},\ and\ \citenamefont
  {Baer}}]{RoiDannyGW}%
  \BibitemOpen
  \bibfield  {author} {\bibinfo {author} {\bibfnamefont {D.}~\bibnamefont
  {Neuhauser}}, \bibinfo {author} {\bibfnamefont {Y.}~\bibnamefont {Gao}},
  \bibinfo {author} {\bibfnamefont {C.}~\bibnamefont {Arntsen}}, \bibinfo
  {author} {\bibfnamefont {C.}~\bibnamefont {Karshenas}}, \bibinfo {author}
  {\bibfnamefont {E.}~\bibnamefont {Rabani}},\ and\ \bibinfo {author}
  {\bibfnamefont {R.}~\bibnamefont {Baer}},\ }\bibfield  {title} {\bibinfo
  {title} {Breaking the theoretical scaling limit for predicting quasiparticle
  energies: The stochastic $gw$ approach},\ }\href
  {https://doi.org/10.1103/PhysRevLett.113.076402} {\bibfield  {journal}
  {\bibinfo  {journal} {Phys. Rev. Lett.}\ }\textbf {\bibinfo {volume} {113}},\
  \bibinfo {pages} {076402} (\bibinfo {year} {2014})}\BibitemShut {NoStop}%
\bibitem [{\citenamefont {Vl{\v{c}}ek}\ \emph {et~al.}(2017)\citenamefont
  {Vl{\v{c}}ek}, \citenamefont {Rabani}, \citenamefont {Neuhauser},\ and\
  \citenamefont {Baer}}]{VojtechStochGW}%
  \BibitemOpen
  \bibfield  {author} {\bibinfo {author} {\bibfnamefont {V.}~\bibnamefont
  {Vl{\v{c}}ek}}, \bibinfo {author} {\bibfnamefont {E.}~\bibnamefont {Rabani}},
  \bibinfo {author} {\bibfnamefont {D.}~\bibnamefont {Neuhauser}},\ and\
  \bibinfo {author} {\bibfnamefont {R.}~\bibnamefont {Baer}},\ }\bibfield
  {title} {\bibinfo {title} {Stochastic gw calculations for molecules},\ }\href
  {https://doi.org/10.1021/acs.jctc.7b00770} {\bibfield  {journal} {\bibinfo
  {journal} {Journal of Chemical Theory and Computation}\ }\textbf {\bibinfo
  {volume} {13}},\ \bibinfo {pages} {4997} (\bibinfo {year} {2017})},\ \bibinfo
  {note} {pMID: 28876912},\ \Eprint
  {https://arxiv.org/abs/https://doi.org/10.1021/acs.jctc.7b00770}
  {https://doi.org/10.1021/acs.jctc.7b00770} \BibitemShut {NoStop}%
\bibitem [{\citenamefont {Vl{\v{c}}ek}\ \emph {et~al.}(2018)\citenamefont
  {Vl{\v{c}}ek}, \citenamefont {Li}, \citenamefont {Baer}, \citenamefont
  {Rabani},\ and\ \citenamefont {Neuhauser}}]{vlcek2018swift}%
  \BibitemOpen
  \bibfield  {author} {\bibinfo {author} {\bibfnamefont {V.}~\bibnamefont
  {Vl{\v{c}}ek}}, \bibinfo {author} {\bibfnamefont {W.}~\bibnamefont {Li}},
  \bibinfo {author} {\bibfnamefont {R.}~\bibnamefont {Baer}}, \bibinfo {author}
  {\bibfnamefont {E.}~\bibnamefont {Rabani}},\ and\ \bibinfo {author}
  {\bibfnamefont {D.}~\bibnamefont {Neuhauser}},\ }\bibfield  {title} {\bibinfo
  {title} {Swift g w beyond 10,000 electrons using sparse stochastic
  compression},\ }\href {https://doi.org/10.1103/PhysRevB.98.075107} {\bibfield
   {journal} {\bibinfo  {journal} {Physical Review B}\ }\textbf {\bibinfo
  {volume} {98}},\ \bibinfo {pages} {075107} (\bibinfo {year}
  {2018})}\BibitemShut {NoStop}%
\bibitem [{\citenamefont {Vl{\v{c}}ek}(2019)}]{StochasticVertexcorrections}%
  \BibitemOpen
  \bibfield  {author} {\bibinfo {author} {\bibfnamefont {V.}~\bibnamefont
  {Vl{\v{c}}ek}},\ }\bibfield  {title} {\bibinfo {title} {Stochastic vertex
  corrections: Linear scaling methods for accurate quasiparticle energies},\
  }\href {https://doi.org/10.1021/acs.jctc.9b00317} {\bibfield  {journal}
  {\bibinfo  {journal} {Journal of Chemical Theory and Computation}\ }\textbf
  {\bibinfo {volume} {15}},\ \bibinfo {pages} {6254} (\bibinfo {year}
  {2019})},\ \bibinfo {note} {pMID: 31557012},\ \Eprint
  {https://arxiv.org/abs/https://doi.org/10.1021/acs.jctc.9b00317}
  {https://doi.org/10.1021/acs.jctc.9b00317} \BibitemShut {NoStop}%
\bibitem [{\citenamefont {Abrams}\ and\ \citenamefont
  {Lloyd}(1997)}]{Abrams1997}%
  \BibitemOpen
  \bibfield  {author} {\bibinfo {author} {\bibfnamefont {D.~S.}\ \bibnamefont
  {Abrams}}\ and\ \bibinfo {author} {\bibfnamefont {S.}~\bibnamefont {Lloyd}},\
  }\bibfield  {title} {\bibinfo {title} {Simulation of many-body fermi systems
  on a universal quantum computer},\ }\href
  {https://doi.org/10.1103/PhysRevLett.79.2586} {\bibfield  {journal} {\bibinfo
   {journal} {Phys. Rev. Lett.}\ }\textbf {\bibinfo {volume} {79}},\ \bibinfo
  {pages} {2586} (\bibinfo {year} {1997})}\BibitemShut {NoStop}%
\bibitem [{\citenamefont {Lloyd}(1996)}]{Lloyd1996}%
  \BibitemOpen
  \bibfield  {author} {\bibinfo {author} {\bibfnamefont {S.}~\bibnamefont
  {Lloyd}},\ }\bibfield  {title} {\bibinfo {title} {{Universal Quantum
  Simulators}},\ }\href {https://doi.org/10.1126/science.273.5278.1073}
  {\bibfield  {journal} {\bibinfo  {journal} {Science}\ }\textbf {\bibinfo
  {volume} {273}},\ \bibinfo {pages} {1073} (\bibinfo {year}
  {1996})}\BibitemShut {NoStop}%
\bibitem [{\citenamefont {Kimmel}\ \emph {et~al.}(2015)\citenamefont {Kimmel},
  \citenamefont {Low},\ and\ \citenamefont {Yoder}}]{kimmel-rpe}%
  \BibitemOpen
  \bibfield  {author} {\bibinfo {author} {\bibfnamefont {S.}~\bibnamefont
  {Kimmel}}, \bibinfo {author} {\bibfnamefont {G.~H.}\ \bibnamefont {Low}},\
  and\ \bibinfo {author} {\bibfnamefont {T.~J.}\ \bibnamefont {Yoder}},\
  }\bibfield  {title} {\bibinfo {title} {Robust calibration of a universal
  single-qubit gate set via robust phase estimation},\ }\href
  {https://doi.org/10.1103/PhysRevA.92.062315} {\bibfield  {journal} {\bibinfo
  {journal} {Phys. Rev. A}\ }\textbf {\bibinfo {volume} {92}},\ \bibinfo
  {pages} {062315} (\bibinfo {year} {2015})}\BibitemShut {NoStop}%
\bibitem [{\citenamefont {Rudinger}\ \emph {et~al.}(2017)\citenamefont
  {Rudinger}, \citenamefont {Kimmel}, \citenamefont {Lobser},\ and\
  \citenamefont {Maunz}}]{kenneth-rpe}%
  \BibitemOpen
  \bibfield  {author} {\bibinfo {author} {\bibfnamefont {K.}~\bibnamefont
  {Rudinger}}, \bibinfo {author} {\bibfnamefont {S.}~\bibnamefont {Kimmel}},
  \bibinfo {author} {\bibfnamefont {D.}~\bibnamefont {Lobser}},\ and\ \bibinfo
  {author} {\bibfnamefont {P.}~\bibnamefont {Maunz}},\ }\bibfield  {title}
  {\bibinfo {title} {Experimental demonstration of a cheap and accurate phase
  estimation},\ }\href {https://doi.org/10.1103/PhysRevLett.118.190502}
  {\bibfield  {journal} {\bibinfo  {journal} {Phys. Rev. Lett.}\ }\textbf
  {\bibinfo {volume} {118}},\ \bibinfo {pages} {190502} (\bibinfo {year}
  {2017})}\BibitemShut {NoStop}%
\bibitem [{\citenamefont {Shen}\ \emph
  {et~al.}(2023{\natexlab{a}})\citenamefont {Shen}, \citenamefont {Klymko},
  \citenamefont {Sud}, \citenamefont {Williams-Young}, \citenamefont {Jong},\
  and\ \citenamefont {Tubman}}]{yizhi-vqpe-krylov}%
  \BibitemOpen
  \bibfield  {author} {\bibinfo {author} {\bibfnamefont {Y.}~\bibnamefont
  {Shen}}, \bibinfo {author} {\bibfnamefont {K.}~\bibnamefont {Klymko}},
  \bibinfo {author} {\bibfnamefont {J.}~\bibnamefont {Sud}}, \bibinfo {author}
  {\bibfnamefont {D.~B.}\ \bibnamefont {Williams-Young}}, \bibinfo {author}
  {\bibfnamefont {W.~A.~d.}\ \bibnamefont {Jong}},\ and\ \bibinfo {author}
  {\bibfnamefont {N.~M.}\ \bibnamefont {Tubman}},\ }\bibfield  {title}
  {\bibinfo {title} {Real-{T}ime {K}rylov {T}heory for {Q}uantum {C}omputing
  {A}lgorithms},\ }\href {https://doi.org/10.22331/q-2023-07-25-1066}
  {\bibfield  {journal} {\bibinfo  {journal} {{Quantum}}\ }\textbf {\bibinfo
  {volume} {7}},\ \bibinfo {pages} {1066} (\bibinfo {year}
  {2023}{\natexlab{a}})}\BibitemShut {NoStop}%
\bibitem [{\citenamefont {Klymko}\ \emph {et~al.}(2022)\citenamefont {Klymko},
  \citenamefont {Mejuto-Zaera}, \citenamefont {Cotton}, \citenamefont
  {Wudarski}, \citenamefont {Urbanek}, \citenamefont {Hait}, \citenamefont
  {Head-Gordon}, \citenamefont {Whaley}, \citenamefont {Moussa}, \citenamefont
  {Wiebe}, \citenamefont {de~Jong},\ and\ \citenamefont
  {Tubman}}]{klymko-vqpe}%
  \BibitemOpen
  \bibfield  {author} {\bibinfo {author} {\bibfnamefont {K.}~\bibnamefont
  {Klymko}}, \bibinfo {author} {\bibfnamefont {C.}~\bibnamefont
  {Mejuto-Zaera}}, \bibinfo {author} {\bibfnamefont {S.~J.}\ \bibnamefont
  {Cotton}}, \bibinfo {author} {\bibfnamefont {F.}~\bibnamefont {Wudarski}},
  \bibinfo {author} {\bibfnamefont {M.}~\bibnamefont {Urbanek}}, \bibinfo
  {author} {\bibfnamefont {D.}~\bibnamefont {Hait}}, \bibinfo {author}
  {\bibfnamefont {M.}~\bibnamefont {Head-Gordon}}, \bibinfo {author}
  {\bibfnamefont {K.~B.}\ \bibnamefont {Whaley}}, \bibinfo {author}
  {\bibfnamefont {J.}~\bibnamefont {Moussa}}, \bibinfo {author} {\bibfnamefont
  {N.}~\bibnamefont {Wiebe}}, \bibinfo {author} {\bibfnamefont {W.~A.}\
  \bibnamefont {de~Jong}},\ and\ \bibinfo {author} {\bibfnamefont {N.~M.}\
  \bibnamefont {Tubman}},\ }\bibfield  {title} {\bibinfo {title} {Real-time
  evolution for ultracompact hamiltonian eigenstates on quantum hardware},\
  }\href {https://doi.org/10.1103/PRXQuantum.3.020323} {\bibfield  {journal}
  {\bibinfo  {journal} {PRX Quantum}\ }\textbf {\bibinfo {volume} {3}},\
  \bibinfo {pages} {020323} (\bibinfo {year} {2022})}\BibitemShut {NoStop}%
\bibitem [{\citenamefont {Shen}\ \emph
  {et~al.}(2023{\natexlab{b}})\citenamefont {Shen}, \citenamefont {Camps},
  \citenamefont {Szasz}, \citenamefont {Darbha}, \citenamefont {Klymko},
  \citenamefont {Williams-Young}, \citenamefont {Tubman},\ and\ \citenamefont
  {Beeumen}}]{shen-vqpe}%
  \BibitemOpen
  \bibfield  {author} {\bibinfo {author} {\bibfnamefont {Y.}~\bibnamefont
  {Shen}}, \bibinfo {author} {\bibfnamefont {D.}~\bibnamefont {Camps}},
  \bibinfo {author} {\bibfnamefont {A.}~\bibnamefont {Szasz}}, \bibinfo
  {author} {\bibfnamefont {S.}~\bibnamefont {Darbha}}, \bibinfo {author}
  {\bibfnamefont {K.}~\bibnamefont {Klymko}}, \bibinfo {author} {\bibfnamefont
  {D.~B.}\ \bibnamefont {Williams-Young}}, \bibinfo {author} {\bibfnamefont
  {N.~M.}\ \bibnamefont {Tubman}},\ and\ \bibinfo {author} {\bibfnamefont
  {R.~V.}\ \bibnamefont {Beeumen}},\ }\href@noop {} {\bibinfo {title}
  {Estimating eigenenergies from quantum dynamics: A unified noise-resilient
  measurement-driven approach}} (\bibinfo {year} {2023}{\natexlab{b}}),\
  \Eprint {https://arxiv.org/abs/2306.01858} {arXiv:2306.01858 [quant-ph]}
  \BibitemShut {NoStop}%
\bibitem [{\citenamefont {Cohn}\ \emph {et~al.}(2021)\citenamefont {Cohn},
  \citenamefont {Motta},\ and\ \citenamefont {Parrish}}]{cohn-qfd}%
  \BibitemOpen
  \bibfield  {author} {\bibinfo {author} {\bibfnamefont {J.}~\bibnamefont
  {Cohn}}, \bibinfo {author} {\bibfnamefont {M.}~\bibnamefont {Motta}},\ and\
  \bibinfo {author} {\bibfnamefont {R.~M.}\ \bibnamefont {Parrish}},\
  }\bibfield  {title} {\bibinfo {title} {Quantum filter diagonalization with
  compressed double-factorized hamiltonians},\ }\href
  {https://doi.org/10.1103/PRXQuantum.2.040352} {\bibfield  {journal} {\bibinfo
   {journal} {PRX Quantum}\ }\textbf {\bibinfo {volume} {2}},\ \bibinfo {pages}
  {040352} (\bibinfo {year} {2021})}\BibitemShut {NoStop}%
\bibitem [{\citenamefont {Stenger}\ \emph {et~al.}(2022)\citenamefont
  {Stenger}, \citenamefont {Ben-Shach}, \citenamefont {Pekker},\ and\
  \citenamefont {Bronn}}]{Stenger2022}%
  \BibitemOpen
  \bibfield  {author} {\bibinfo {author} {\bibfnamefont {J.~P.}\ \bibnamefont
  {Stenger}}, \bibinfo {author} {\bibfnamefont {G.}~\bibnamefont {Ben-Shach}},
  \bibinfo {author} {\bibfnamefont {D.}~\bibnamefont {Pekker}},\ and\ \bibinfo
  {author} {\bibfnamefont {N.~T.}\ \bibnamefont {Bronn}},\ }\bibfield  {title}
  {\bibinfo {title} {{Simulating spectroscopy experiments with a
  superconducting quantum computer}},\ }\href
  {https://doi.org/10.1103/PhysRevResearch.4.043106} {\bibfield  {journal}
  {\bibinfo  {journal} {Physical Review Research}\ }\textbf {\bibinfo {volume}
  {4}},\ \bibinfo {pages} {043106} (\bibinfo {year} {2022})}\BibitemShut
  {NoStop}%
\bibitem [{\citenamefont {C{\'{o}}rcoles}\ \emph {et~al.}(2021)\citenamefont
  {C{\'{o}}rcoles}, \citenamefont {Takita}, \citenamefont {Inoue},
  \citenamefont {Lekuch}, \citenamefont {Minev}, \citenamefont {Chow},\ and\
  \citenamefont {Gambetta}}]{Corcoles2021}%
  \BibitemOpen
  \bibfield  {author} {\bibinfo {author} {\bibfnamefont {A.~D.}\ \bibnamefont
  {C{\'{o}}rcoles}}, \bibinfo {author} {\bibfnamefont {M.}~\bibnamefont
  {Takita}}, \bibinfo {author} {\bibfnamefont {K.}~\bibnamefont {Inoue}},
  \bibinfo {author} {\bibfnamefont {S.}~\bibnamefont {Lekuch}}, \bibinfo
  {author} {\bibfnamefont {Z.~K.}\ \bibnamefont {Minev}}, \bibinfo {author}
  {\bibfnamefont {J.~M.}\ \bibnamefont {Chow}},\ and\ \bibinfo {author}
  {\bibfnamefont {J.~M.}\ \bibnamefont {Gambetta}},\ }\bibfield  {title}
  {\bibinfo {title} {{Exploiting dynamic quantum circuits in a quantum
  algorithm with superconducting qubits}},\ }\href
  {https://doi.org/10.1103/PhysRevLett.127.100501} {\bibfield  {journal}
  {\bibinfo  {journal} {Physical Review Letters}\ }\textbf {\bibinfo {volume}
  {127}},\ \bibinfo {pages} {100501} (\bibinfo {year} {2021})}\BibitemShut
  {NoStop}%
\bibitem [{\citenamefont {Somma}\ \emph {et~al.}(2002)\citenamefont {Somma},
  \citenamefont {Ortiz}, \citenamefont {Gubernatis}, \citenamefont {Knill},\
  and\ \citenamefont {Laflamme}}]{Somma2002}%
  \BibitemOpen
  \bibfield  {author} {\bibinfo {author} {\bibfnamefont {R.}~\bibnamefont
  {Somma}}, \bibinfo {author} {\bibfnamefont {G.}~\bibnamefont {Ortiz}},
  \bibinfo {author} {\bibfnamefont {J.~E.}\ \bibnamefont {Gubernatis}},
  \bibinfo {author} {\bibfnamefont {E.}~\bibnamefont {Knill}},\ and\ \bibinfo
  {author} {\bibfnamefont {R.}~\bibnamefont {Laflamme}},\ }\bibfield  {title}
  {\bibinfo {title} {{Simulating physical phenomena by quantum networks}},\
  }\href {https://doi.org/10.1103/PhysRevA.65.042323} {\bibfield  {journal}
  {\bibinfo  {journal} {Physical Review A}\ }\textbf {\bibinfo {volume} {65}},\
  \bibinfo {pages} {042323} (\bibinfo {year} {2002})}\BibitemShut {NoStop}%
\bibitem [{\citenamefont {Somma}(2019)}]{Somma2019}%
  \BibitemOpen
  \bibfield  {author} {\bibinfo {author} {\bibfnamefont {R.~D.}\ \bibnamefont
  {Somma}},\ }\bibfield  {title} {\bibinfo {title} {{Quantum eigenvalue
  estimation via time series analysis}},\ }\href
  {https://iopscience.iop.org/article/10.1088/1367-2630/ab5c60} {\bibfield
  {journal} {\bibinfo  {journal} {New Journal of Physics}\ }\textbf {\bibinfo
  {volume} {21}},\ \bibinfo {pages} {123025} (\bibinfo {year}
  {2019})}\BibitemShut {NoStop}%
\bibitem [{\citenamefont {Hutchinson}(1989)}]{HutchinsonTrace}%
  \BibitemOpen
  \bibfield  {author} {\bibinfo {author} {\bibfnamefont {M.}~\bibnamefont
  {Hutchinson}},\ }\bibfield  {title} {\bibinfo {title} {A stochastic estimator
  of the trace of the influence matrix for laplacian smoothing splines},\
  }\href {https://doi.org/10.1080/03610918908812806} {\bibfield  {journal}
  {\bibinfo  {journal} {Communications in Statistics - Simulation and
  Computation}\ }\textbf {\bibinfo {volume} {18}},\ \bibinfo {pages} {1059}
  (\bibinfo {year} {1989})},\ \Eprint
  {https://arxiv.org/abs/https://doi.org/10.1080/03610918908812806}
  {https://doi.org/10.1080/03610918908812806} \BibitemShut {NoStop}%
\bibitem [{\citenamefont {Neuhauser}(1990)}]{NeuhauserFD}%
  \BibitemOpen
  \bibfield  {author} {\bibinfo {author} {\bibfnamefont {D.}~\bibnamefont
  {Neuhauser}},\ }\bibfield  {title} {\bibinfo {title} {{Bound state
  eigenfunctions from wave packets: Time→energy resolution}},\ }\href
  {https://doi.org/10.1063/1.458900} {\bibfield  {journal} {\bibinfo  {journal}
  {The Journal of Chemical Physics}\ }\textbf {\bibinfo {volume} {93}},\
  \bibinfo {pages} {2611} (\bibinfo {year} {1990})},\ \Eprint
  {https://arxiv.org/abs/https://pubs.aip.org/aip/jcp/article-pdf/93/4/2611/11340586/2611\_1\_online.pdf}
  {https://pubs.aip.org/aip/jcp/article-pdf/93/4/2611/11340586/2611\_1\_online.pdf}
  \BibitemShut {NoStop}%
\bibitem [{\citenamefont {Wall}\ and\ \citenamefont
  {Neuhauser}(1995)}]{NeuhauserFD2}%
  \BibitemOpen
  \bibfield  {author} {\bibinfo {author} {\bibfnamefont {M.~R.}\ \bibnamefont
  {Wall}}\ and\ \bibinfo {author} {\bibfnamefont {D.}~\bibnamefont
  {Neuhauser}},\ }\bibfield  {title} {\bibinfo {title} {{Extraction, through
  filter‐diagonalization, of general quantum eigenvalues or classical normal
  mode frequencies from a small number of residues or a short‐time segment of
  a signal. I. Theory and application to a quantum‐dynamics model}},\ }\href
  {https://doi.org/10.1063/1.468999} {\bibfield  {journal} {\bibinfo  {journal}
  {The Journal of Chemical Physics}\ }\textbf {\bibinfo {volume} {102}},\
  \bibinfo {pages} {8011} (\bibinfo {year} {1995})}\BibitemShut {NoStop}%
\bibitem [{\citenamefont {Mandelshtam}\ and\ \citenamefont
  {Taylor}(1997)}]{Mandelshtam}%
  \BibitemOpen
  \bibfield  {author} {\bibinfo {author} {\bibfnamefont {V.~A.}\ \bibnamefont
  {Mandelshtam}}\ and\ \bibinfo {author} {\bibfnamefont {H.~S.}\ \bibnamefont
  {Taylor}},\ }\bibfield  {title} {\bibinfo {title} {{A low-storage filter
  diagonalization method for quantum eigenenergy calculation or for spectral
  analysis of time signals}},\ }\href {https://doi.org/10.1063/1.473554}
  {\bibfield  {journal} {\bibinfo  {journal} {The Journal of Chemical Physics}\
  }\textbf {\bibinfo {volume} {106}},\ \bibinfo {pages} {5085} (\bibinfo {year}
  {1997})},\ \Eprint
  {https://arxiv.org/abs/https://pubs.aip.org/aip/jcp/article-pdf/106/12/5085/10781975/5085\_1\_online.pdf}
  {https://pubs.aip.org/aip/jcp/article-pdf/106/12/5085/10781975/5085\_1\_online.pdf}
  \BibitemShut {NoStop}%
\bibitem [{\citenamefont {Preskill}(2018)}]{Preskill2018}%
  \BibitemOpen
  \bibfield  {author} {\bibinfo {author} {\bibfnamefont {J.}~\bibnamefont
  {Preskill}},\ }\bibfield  {title} {\bibinfo {title} {{Quantum computing in
  the NISQ era and beyond}},\ }\href {https://doi.org/10.22331/q-2018-08-06-79}
  {\bibfield  {journal} {\bibinfo  {journal} {Quantum}\ }\textbf {\bibinfo
  {volume} {2}},\ \bibinfo {pages} {79} (\bibinfo {year} {2018})}\BibitemShut
  {NoStop}%
\bibitem [{\citenamefont {Chamberland}\ \emph {et~al.}(2020)\citenamefont
  {Chamberland}, \citenamefont {Zhu}, \citenamefont {Yoder}, \citenamefont
  {Hertzberg},\ and\ \citenamefont {Cross}}]{Chamberland2019}%
  \BibitemOpen
  \bibfield  {author} {\bibinfo {author} {\bibfnamefont {C.}~\bibnamefont
  {Chamberland}}, \bibinfo {author} {\bibfnamefont {G.}~\bibnamefont {Zhu}},
  \bibinfo {author} {\bibfnamefont {T.~J.}\ \bibnamefont {Yoder}}, \bibinfo
  {author} {\bibfnamefont {J.~B.}\ \bibnamefont {Hertzberg}},\ and\ \bibinfo
  {author} {\bibfnamefont {A.~W.}\ \bibnamefont {Cross}},\ }\bibfield  {title}
  {\bibinfo {title} {{Topological and subsystem codes on low-degree graphs with
  flag qubits}},\ }\href {https://doi.org/10.1103/PhysRevX.10.011022}
  {\bibfield  {journal} {\bibinfo  {journal} {Phys. Rev. X}\ }\textbf {\bibinfo
  {volume} {10}},\ \bibinfo {pages} {011022} (\bibinfo {year}
  {2020})}\BibitemShut {NoStop}%
\bibitem [{\citenamefont {McKay}\ \emph {et~al.}(2017)\citenamefont {McKay},
  \citenamefont {Wood}, \citenamefont {Sheldon}, \citenamefont {Chow},\ and\
  \citenamefont {Gambetta}}]{McKay2017}%
  \BibitemOpen
  \bibfield  {author} {\bibinfo {author} {\bibfnamefont {D.~C.}\ \bibnamefont
  {McKay}}, \bibinfo {author} {\bibfnamefont {C.~J.}\ \bibnamefont {Wood}},
  \bibinfo {author} {\bibfnamefont {S.}~\bibnamefont {Sheldon}}, \bibinfo
  {author} {\bibfnamefont {J.~M.}\ \bibnamefont {Chow}},\ and\ \bibinfo
  {author} {\bibfnamefont {J.~M.}\ \bibnamefont {Gambetta}},\ }\bibfield
  {title} {\bibinfo {title} {Efficient $z$ gates for quantum computing},\
  }\href {https://doi.org/10.1103/PhysRevA.96.022330} {\bibfield  {journal}
  {\bibinfo  {journal} {Phys. Rev. A}\ }\textbf {\bibinfo {volume} {96}},\
  \bibinfo {pages} {022330} (\bibinfo {year} {2017})}\BibitemShut {NoStop}%
\bibitem [{\citenamefont {Stenger}\ \emph {et~al.}(2021)\citenamefont
  {Stenger}, \citenamefont {Bronn}, \citenamefont {Egger},\ and\ \citenamefont
  {Pekker}}]{Stenger2021}%
  \BibitemOpen
  \bibfield  {author} {\bibinfo {author} {\bibfnamefont {J.~P.~T.}\
  \bibnamefont {Stenger}}, \bibinfo {author} {\bibfnamefont {N.~T.}\
  \bibnamefont {Bronn}}, \bibinfo {author} {\bibfnamefont {D.~J.}\ \bibnamefont
  {Egger}},\ and\ \bibinfo {author} {\bibfnamefont {D.}~\bibnamefont
  {Pekker}},\ }\bibfield  {title} {\bibinfo {title} {Simulating the dynamics of
  braiding of majorana zero modes using an ibm quantum computer},\ }\href
  {https://doi.org/10.1103/PhysRevResearch.3.033171} {\bibfield  {journal}
  {\bibinfo  {journal} {Phys. Rev. Res.}\ }\textbf {\bibinfo {volume} {3}},\
  \bibinfo {pages} {033171} (\bibinfo {year} {2021})}\BibitemShut {NoStop}%
\bibitem [{\citenamefont {Shende}\ \emph {et~al.}(2006)\citenamefont {Shende},
  \citenamefont {Bullock},\ and\ \citenamefont {Markov}}]{shende-2006}%
  \BibitemOpen
  \bibfield  {author} {\bibinfo {author} {\bibfnamefont {V.}~\bibnamefont
  {Shende}}, \bibinfo {author} {\bibfnamefont {S.}~\bibnamefont {Bullock}},\
  and\ \bibinfo {author} {\bibfnamefont {I.}~\bibnamefont {Markov}},\
  }\bibfield  {title} {\bibinfo {title} {Synthesis of quantum-logic circuits},\
  }\href {https://doi.org/10.1109/TCAD.2005.855930} {\bibfield  {journal}
  {\bibinfo  {journal} {IEEE Transactions on Computer-Aided Design of
  Integrated Circuits and Systems}\ }\textbf {\bibinfo {volume} {25}},\
  \bibinfo {pages} {1000} (\bibinfo {year} {2006})}\BibitemShut {NoStop}%
\bibitem [{\citenamefont {Li}\ \emph {et~al.}(2019)\citenamefont {Li},
  \citenamefont {Ding},\ and\ \citenamefont {Xie}}]{sabre-swap}%
  \BibitemOpen
  \bibfield  {author} {\bibinfo {author} {\bibfnamefont {G.}~\bibnamefont
  {Li}}, \bibinfo {author} {\bibfnamefont {Y.}~\bibnamefont {Ding}},\ and\
  \bibinfo {author} {\bibfnamefont {Y.}~\bibnamefont {Xie}},\ }\href@noop {}
  {\bibinfo {title} {Tackling the qubit mapping problem for nisq-era quantum
  devices}} (\bibinfo {year} {2019}),\ \Eprint
  {https://arxiv.org/abs/1809.02573} {arXiv:1809.02573 [cs.ET]} \BibitemShut
  {NoStop}%
\bibitem [{\citenamefont {Nation}\ \emph {et~al.}(2021)\citenamefont {Nation},
  \citenamefont {Kang}, \citenamefont {Sundaresan},\ and\ \citenamefont
  {Gambetta}}]{Nation2021}%
  \BibitemOpen
  \bibfield  {author} {\bibinfo {author} {\bibfnamefont {P.~D.}\ \bibnamefont
  {Nation}}, \bibinfo {author} {\bibfnamefont {H.}~\bibnamefont {Kang}},
  \bibinfo {author} {\bibfnamefont {N.~M.}\ \bibnamefont {Sundaresan}},\ and\
  \bibinfo {author} {\bibfnamefont {J.~M.}\ \bibnamefont {Gambetta}},\
  }\bibfield  {title} {\bibinfo {title} {{Scalable mitigation of measurement
  errors on quantum computers}},\ }\href
  {https://doi.org/10.1103/PRXQuantum.2.040326} {\bibfield  {journal} {\bibinfo
   {journal} {PRX Quantum}\ }\textbf {\bibinfo {volume} {2}},\ \bibinfo {pages}
  {040326} (\bibinfo {year} {2021})}\BibitemShut {NoStop}%
\bibitem [{\citenamefont {Kemper}\ \emph {et~al.}(2023)\citenamefont {Kemper},
  \citenamefont {Yang},\ and\ \citenamefont {Gull}}]{kemper2023positive}%
  \BibitemOpen
  \bibfield  {author} {\bibinfo {author} {\bibfnamefont {A.~F.}\ \bibnamefont
  {Kemper}}, \bibinfo {author} {\bibfnamefont {C.}~\bibnamefont {Yang}},\ and\
  \bibinfo {author} {\bibfnamefont {E.}~\bibnamefont {Gull}},\ }\bibfield
  {title} {\bibinfo {title} {On the positive definiteness of response functions
  in the time domain},\ }\href@noop {} {\bibfield  {journal} {\bibinfo
  {journal} {arXiv preprint arXiv:2309.02566}\ } (\bibinfo {year}
  {2023})}\BibitemShut {NoStop}%
\bibitem [{\citenamefont {Schmid}(2010)}]{DMD0}%
  \BibitemOpen
  \bibfield  {author} {\bibinfo {author} {\bibfnamefont {P.~J.}\ \bibnamefont
  {Schmid}},\ }\bibfield  {title} {\bibinfo {title} {Dynamic mode decomposition
  of numerical and experimental data},\ }\href
  {https://www.cambridge.org/core/journals/journal-of-fluid-mechanics/article/dynamic-mode-decomposition-of-numerical-and-experimental-data/AA4C763B525515AD4521A6CC5E10DBD4}
  {\bibfield  {journal} {\bibinfo  {journal} {J. Fluid Mech.}\ }\textbf
  {\bibinfo {volume} {656}},\ \bibinfo {pages} {5} (\bibinfo {year}
  {2010})}\BibitemShut {NoStop}%
\bibitem [{\citenamefont {Schmid}\ \emph {et~al.}(2011)\citenamefont {Schmid},
  \citenamefont {Li}, \citenamefont {Juniper},\ and\ \citenamefont
  {Pust}}]{schmid2011applications}%
  \BibitemOpen
  \bibfield  {author} {\bibinfo {author} {\bibfnamefont {P.~J.}\ \bibnamefont
  {Schmid}}, \bibinfo {author} {\bibfnamefont {L.}~\bibnamefont {Li}}, \bibinfo
  {author} {\bibfnamefont {M.~P.}\ \bibnamefont {Juniper}},\ and\ \bibinfo
  {author} {\bibfnamefont {O.}~\bibnamefont {Pust}},\ }\bibfield  {title}
  {\bibinfo {title} {Applications of the dynamic mode decomposition},\ }\href
  {https://link.springer.com/article/10.1007/s00162-010-0203-9} {\bibfield
  {journal} {\bibinfo  {journal} {Theor. Comput. Fluid Dyn.}\ }\textbf
  {\bibinfo {volume} {25}},\ \bibinfo {pages} {249} (\bibinfo {year}
  {2011})}\BibitemShut {NoStop}%
\bibitem [{\citenamefont {Tu}\ \emph {et~al.}(2014)\citenamefont {Tu},
  \citenamefont {Rowley}, \citenamefont {Luchtenburg}, \citenamefont
  {Brunton},\ and\ \citenamefont {Kutz}}]{TuRowley}%
  \BibitemOpen
  \bibfield  {author} {\bibinfo {author} {\bibfnamefont {J.~H.}\ \bibnamefont
  {Tu}}, \bibinfo {author} {\bibfnamefont {C.~W.}\ \bibnamefont {Rowley}},
  \bibinfo {author} {\bibfnamefont {D.~M.}\ \bibnamefont {Luchtenburg}},
  \bibinfo {author} {\bibfnamefont {S.~L.}\ \bibnamefont {Brunton}},\ and\
  \bibinfo {author} {\bibfnamefont {J.~N.}\ \bibnamefont {Kutz}},\ }\bibfield
  {title} {\bibinfo {title} {On dynamic mode decomposition: Theory and
  applications},\ }\href
  {https://www.aimsciences.org/article/doi/10.3934/jcd.2014.1.391} {\bibfield
  {journal} {\bibinfo  {journal} {J. Comput. Dyn.}\ }\textbf {\bibinfo {volume}
  {1}},\ \bibinfo {pages} {391} (\bibinfo {year} {2014})}\BibitemShut {NoStop}%
\bibitem [{\citenamefont {Reeves}\ \emph {et~al.}(2023)\citenamefont {Reeves},
  \citenamefont {Yin}, \citenamefont {Zhu}, \citenamefont {Ibrahim},
  \citenamefont {Yang},\ and\ \citenamefont {Vl\ifmmode~\check{c}\else
  \v{c}\fi{}ek}}]{DMDVV}%
  \BibitemOpen
  \bibfield  {author} {\bibinfo {author} {\bibfnamefont {C.~C.}\ \bibnamefont
  {Reeves}}, \bibinfo {author} {\bibfnamefont {J.}~\bibnamefont {Yin}},
  \bibinfo {author} {\bibfnamefont {Y.}~\bibnamefont {Zhu}}, \bibinfo {author}
  {\bibfnamefont {K.~Z.}\ \bibnamefont {Ibrahim}}, \bibinfo {author}
  {\bibfnamefont {C.}~\bibnamefont {Yang}},\ and\ \bibinfo {author}
  {\bibfnamefont {V.~c.~v.}\ \bibnamefont {Vl\ifmmode~\check{c}\else
  \v{c}\fi{}ek}},\ }\bibfield  {title} {\bibinfo {title} {Dynamic mode
  decomposition for extrapolating nonequilibrium green's-function dynamics},\
  }\href {https://doi.org/10.1103/PhysRevB.107.075107} {\bibfield  {journal}
  {\bibinfo  {journal} {Phys. Rev. B}\ }\textbf {\bibinfo {volume} {107}},\
  \bibinfo {pages} {075107} (\bibinfo {year} {2023})}\BibitemShut {NoStop}%
\bibitem [{\citenamefont {Yin}\ \emph {et~al.}(2021)\citenamefont {Yin},
  \citenamefont {Chan}, \citenamefont {da~Jornada}, \citenamefont {Qiu},
  \citenamefont {Yang},\ and\ \citenamefont {Louie}}]{DMDdiag}%
  \BibitemOpen
  \bibfield  {author} {\bibinfo {author} {\bibfnamefont {J.}~\bibnamefont
  {Yin}}, \bibinfo {author} {\bibfnamefont {Y.-h.}\ \bibnamefont {Chan}},
  \bibinfo {author} {\bibfnamefont {F.}~\bibnamefont {da~Jornada}}, \bibinfo
  {author} {\bibfnamefont {D.}~\bibnamefont {Qiu}}, \bibinfo {author}
  {\bibfnamefont {C.}~\bibnamefont {Yang}},\ and\ \bibinfo {author}
  {\bibfnamefont {S.~G.}\ \bibnamefont {Louie}},\ }\bibfield  {title} {\bibinfo
  {title} {Analyzing and predicting non-equilibrium many-body dynamics via
  dynamic mode decomposition},\ }\href {https://arxiv.org/abs/2107.09635}
  {\bibfield  {journal} {\bibinfo  {journal} {arXiv preprint arXiv:2107.09635}\
  } (\bibinfo {year} {2021})}\BibitemShut {NoStop}%
\bibitem [{\citenamefont {Mejía}\ \emph {et~al.}(2023)\citenamefont {Mejía},
  \citenamefont {Yin}, \citenamefont {Reichman}, \citenamefont {Baer},
  \citenamefont {Yang},\ and\ \citenamefont {Rabani}}]{mejía2023stochastic}%
  \BibitemOpen
  \bibfield  {author} {\bibinfo {author} {\bibfnamefont {L.}~\bibnamefont
  {Mejía}}, \bibinfo {author} {\bibfnamefont {J.}~\bibnamefont {Yin}},
  \bibinfo {author} {\bibfnamefont {D.~R.}\ \bibnamefont {Reichman}}, \bibinfo
  {author} {\bibfnamefont {R.}~\bibnamefont {Baer}}, \bibinfo {author}
  {\bibfnamefont {C.}~\bibnamefont {Yang}},\ and\ \bibinfo {author}
  {\bibfnamefont {E.}~\bibnamefont {Rabani}},\ }\href@noop {} {\bibinfo {title}
  {Stochastic real-time second-order green's function theory for neutral
  excitations in molecules and nanostructures}} (\bibinfo {year} {2023}),\
  \Eprint {https://arxiv.org/abs/2303.06874} {arXiv:2303.06874
  [physics.chem-ph]} \BibitemShut {NoStop}%
\bibitem [{\citenamefont {Bassi}\ \emph {et~al.}(2023)\citenamefont {Bassi},
  \citenamefont {Zhu}, \citenamefont {Liang}, \citenamefont {Yin},
  \citenamefont {Reeves}, \citenamefont {Vlcek},\ and\ \citenamefont
  {Yang}}]{bassi2023learning}%
  \BibitemOpen
  \bibfield  {author} {\bibinfo {author} {\bibfnamefont {H.}~\bibnamefont
  {Bassi}}, \bibinfo {author} {\bibfnamefont {Y.}~\bibnamefont {Zhu}}, \bibinfo
  {author} {\bibfnamefont {S.}~\bibnamefont {Liang}}, \bibinfo {author}
  {\bibfnamefont {J.}~\bibnamefont {Yin}}, \bibinfo {author} {\bibfnamefont
  {C.~C.}\ \bibnamefont {Reeves}}, \bibinfo {author} {\bibfnamefont
  {V.}~\bibnamefont {Vlcek}},\ and\ \bibinfo {author} {\bibfnamefont
  {C.}~\bibnamefont {Yang}},\ }\href@noop {} {\bibinfo {title} {Learning
  nonlinear integral operators via recurrent neural networks and its
  application in solving integro-differential equations}} (\bibinfo {year}
  {2023}),\ \Eprint {https://arxiv.org/abs/2310.09434} {arXiv:2310.09434
  [cs.LG]} \BibitemShut {NoStop}%
\bibitem [{\citenamefont {Temme}\ \emph {et~al.}(2017)\citenamefont {Temme},
  \citenamefont {Bravyi},\ and\ \citenamefont {Gambetta}}]{Temme2017}%
  \BibitemOpen
  \bibfield  {author} {\bibinfo {author} {\bibfnamefont {K.}~\bibnamefont
  {Temme}}, \bibinfo {author} {\bibfnamefont {S.}~\bibnamefont {Bravyi}},\ and\
  \bibinfo {author} {\bibfnamefont {J.~M.}\ \bibnamefont {Gambetta}},\
  }\bibfield  {title} {\bibinfo {title} {Error mitigation for short-depth
  quantum circuits},\ }\href {https://doi.org/10.1103/PhysRevLett.119.180509}
  {\bibfield  {journal} {\bibinfo  {journal} {Phys. Rev. Lett.}\ }\textbf
  {\bibinfo {volume} {119}},\ \bibinfo {pages} {180509} (\bibinfo {year}
  {2017})}\BibitemShut {NoStop}%
\bibitem [{\citenamefont {Sung}\ \emph {et~al.}(2022)\citenamefont {Sung},
  \citenamefont {Rancic}, \citenamefont {Lanes},\ and\ \citenamefont
  {Bronn}}]{Sung2022}%
  \BibitemOpen
  \bibfield  {author} {\bibinfo {author} {\bibfnamefont {K.~J.}\ \bibnamefont
  {Sung}}, \bibinfo {author} {\bibfnamefont {M.~J.}\ \bibnamefont {Rancic}},
  \bibinfo {author} {\bibfnamefont {O.~T.}\ \bibnamefont {Lanes}},\ and\
  \bibinfo {author} {\bibfnamefont {N.~T.}\ \bibnamefont {Bronn}},\ }\href@noop
  {} {\bibinfo {title} {Preparing majorana zero modes on a noisy quantum
  processor}} (\bibinfo {year} {2022}),\ \Eprint
  {https://arxiv.org/abs/arXiv:2206.00563} {arXiv:arXiv:2206.00563}
  \BibitemShut {NoStop}%
\bibitem [{\citenamefont {Kim}\ \emph {et~al.}(2023)\citenamefont {Kim},
  \citenamefont {Eddins}, \citenamefont {Anand}, \citenamefont {Wei},
  \citenamefont {van~den Berg}, \citenamefont {Rosenblatt}, \citenamefont
  {Nayfeh}, \citenamefont {Wu}, \citenamefont {Zaletel}, \citenamefont
  {Temme},\ and\ \citenamefont {Kandala}}]{Kim2023}%
  \BibitemOpen
  \bibfield  {author} {\bibinfo {author} {\bibfnamefont {Y.}~\bibnamefont
  {Kim}}, \bibinfo {author} {\bibfnamefont {A.}~\bibnamefont {Eddins}},
  \bibinfo {author} {\bibfnamefont {S.}~\bibnamefont {Anand}}, \bibinfo
  {author} {\bibfnamefont {K.~X.}\ \bibnamefont {Wei}}, \bibinfo {author}
  {\bibfnamefont {E.}~\bibnamefont {van~den Berg}}, \bibinfo {author}
  {\bibfnamefont {S.}~\bibnamefont {Rosenblatt}}, \bibinfo {author}
  {\bibfnamefont {H.}~\bibnamefont {Nayfeh}}, \bibinfo {author} {\bibfnamefont
  {Y.}~\bibnamefont {Wu}}, \bibinfo {author} {\bibfnamefont {M.}~\bibnamefont
  {Zaletel}}, \bibinfo {author} {\bibfnamefont {K.}~\bibnamefont {Temme}},\
  and\ \bibinfo {author} {\bibfnamefont {A.}~\bibnamefont {Kandala}},\
  }\bibfield  {title} {\bibinfo {title} {{Evidence for the utility of quantum
  computing before fault tolerance}},\ }\href
  {https://doi.org/10.1038/s41586-023-06096-3} {\bibfield  {journal} {\bibinfo
  {journal} {Nature}\ }\textbf {\bibinfo {volume} {618}},\ \bibinfo {pages}
  {500} (\bibinfo {year} {2023})}\BibitemShut {NoStop}%
\bibitem [{\citenamefont {van~den Berg}\ \emph {et~al.}(2023)\citenamefont
  {van~den Berg}, \citenamefont {Minev}, \citenamefont {Kandala},\ and\
  \citenamefont {Temme}}]{VandenBerg2023}%
  \BibitemOpen
  \bibfield  {author} {\bibinfo {author} {\bibfnamefont {E.}~\bibnamefont
  {van~den Berg}}, \bibinfo {author} {\bibfnamefont {Z.~K.}\ \bibnamefont
  {Minev}}, \bibinfo {author} {\bibfnamefont {A.}~\bibnamefont {Kandala}},\
  and\ \bibinfo {author} {\bibfnamefont {K.}~\bibnamefont {Temme}},\ }\bibfield
   {title} {\bibinfo {title} {{Probabilistic error cancellation with sparse
  Pauli–Lindblad models on noisy quantum processors}},\ }\href
  {https://doi.org/10.1038/s41567-023-02042-2} {\bibfield  {journal} {\bibinfo
  {journal} {Nature Physics}\ }\textbf {\bibinfo {volume} {19}},\ \bibinfo
  {pages} {1116} (\bibinfo {year} {2023})}\BibitemShut {NoStop}%
\bibitem [{\citenamefont {Majumdar}\ \emph {et~al.}(2023)\citenamefont
  {Majumdar}, \citenamefont {Rivero}, \citenamefont {Metz}, \citenamefont
  {Hasan},\ and\ \citenamefont {Wang}}]{Majumdar2023}%
  \BibitemOpen
  \bibfield  {author} {\bibinfo {author} {\bibfnamefont {R.}~\bibnamefont
  {Majumdar}}, \bibinfo {author} {\bibfnamefont {P.}~\bibnamefont {Rivero}},
  \bibinfo {author} {\bibfnamefont {F.}~\bibnamefont {Metz}}, \bibinfo {author}
  {\bibfnamefont {A.}~\bibnamefont {Hasan}},\ and\ \bibinfo {author}
  {\bibfnamefont {D.~S.}\ \bibnamefont {Wang}},\ }\href
  {http://arxiv.org/abs/2307.05203} {\bibinfo {title} {{Best practices for
  quantum error mitigation with digital zero-noise extrapolation}}} (\bibinfo
  {year} {2023}),\ \Eprint {https://arxiv.org/abs/2307.05203}
  {arXiv:2307.05203} \BibitemShut {NoStop}%
\bibitem [{\citenamefont {Mitarai}\ and\ \citenamefont
  {Fujii}(2019)}]{Mitarai2019}%
  \BibitemOpen
  \bibfield  {author} {\bibinfo {author} {\bibfnamefont {K.}~\bibnamefont
  {Mitarai}}\ and\ \bibinfo {author} {\bibfnamefont {K.}~\bibnamefont
  {Fujii}},\ }\bibfield  {title} {\bibinfo {title} {Methodology for replacing
  indirect measurements with direct measurements},\ }\href
  {https://doi.org/10.1103/PhysRevResearch.1.013006} {\bibfield  {journal}
  {\bibinfo  {journal} {Phys. Rev. Res.}\ }\textbf {\bibinfo {volume} {1}},\
  \bibinfo {pages} {013006} (\bibinfo {year} {2019})}\BibitemShut {NoStop}%
\end{thebibliography}%

\end{document}